\documentclass[%
reprint,
superscriptaddress,
 amsmath,amssymb,
 aps,
]{revtex4-2}

\usepackage{graphicx}%
\usepackage{multirow}%
\usepackage{amsmath,amssymb,amsfonts}%
\usepackage{amsthm}%
\usepackage{mathrsfs}%
\usepackage[title]{appendix}%
\usepackage{xcolor}%
\usepackage{textcomp}%
\usepackage{booktabs}%
\usepackage{listings}%
\usepackage{indentfirst}
\usepackage{float}
\usepackage{hyperref}
\usepackage[all]{hypcap}
\hypersetup{
    colorlinks=true,
    linkcolor=blue,
    filecolor=blue,      
    urlcolor=blue,
    citecolor=blue
    }

\usepackage{soul}

\newcommand{\ket}[1]{\left|{#1}\right\rangle}

\newcommand{\beq}{\begin{equation}}
\newcommand{\eeq}{\end{equation}}
\newcommand{\beqa}{\begin{eqnarray}}
\newcommand{\eeqa}{\end{eqnarray}}

\usepackage{titlesec}
\titleformat*{\section}{\centering\footnotesize\bfseries\uppercase}

\usepackage[paperwidth=210mm,paperheight=297mm,centering,hmargin=1.7cm,vmargin=1.5cm]{geometry}

\begin{document}

\title{\large \textbf{Fluctuation thermometry of an atom-resolved quantum gas:\\Beyond the fluctuation-dissipation theorem}}

\author{Maxime Dixmerias}
\thanks{These authors contributed equally to this work.}
\affiliation{Laboratoire Kastler Brossel, ENS-Universit\'{e} PSL, CNRS, Sorbonne Universit\'{e}, Coll\`{e}ge de France, 24 rue Lhomond, 75005, Paris, France}
\author{Joris Verstraten}
\thanks{These authors contributed equally to this work.}
\affiliation{Laboratoire Kastler Brossel, ENS-Universit\'{e} PSL, CNRS, Sorbonne Universit\'{e}, Coll\`{e}ge de France, 24 rue Lhomond, 75005, Paris, France}
\author{Cyprien Daix}
\affiliation{Laboratoire Kastler Brossel, ENS-Universit\'{e} PSL, CNRS, Sorbonne Universit\'{e}, Coll\`{e}ge de France, 24 rue Lhomond, 75005, Paris, France}
\author{Bruno Peaudecerf}
\affiliation{Laboratoire Collisions Agr\'egats R\'eactivit\'e, UMR 5589, FERMI, UT3, Universit\'e de Toulouse, CNRS, 118 Route de Narbonne, 31062, Toulouse CEDEX 09, France}
\author{Tim de Jongh}
\affiliation{Laboratoire Kastler Brossel, ENS-Universit\'{e} PSL, CNRS, Sorbonne Universit\'{e}, Coll\`{e}ge de France, 24 rue Lhomond, 75005, Paris, France}
\author{Tarik Yefsah}
\affiliation{Laboratoire Kastler Brossel, ENS-Universit\'{e} PSL, CNRS, Sorbonne Universit\'{e}, Coll\`{e}ge de France, 24 rue Lhomond, 75005, Paris, France}

\date{February 7, 2025}

\begin{abstract}
 \quad Thermometry is essential for studying many-body physics with ultracold atoms. Accurately measuring low temperatures in these systems, however, remains a significant challenge due to the absence of a universal thermometer. Most widely applicable methods, such as fitting of in-situ density profiles or standard fluctuation thermometry, are limited by the requirement of global thermal equilibrium and inapplicability to homogeneous systems. In this work, we introduce a novel in-situ thermometry for quantum gases, leveraging single-atom resolved measurements via quantum gas microscopy, and demonstrate it on an ideal Fermi gas. By analyzing number fluctuations in probe volumes with approximately one atom on average, we extract both global and local temperatures over a broad dynamic range. Unlike traditional fluctuation thermometry, our method does not rely on the fluctuation-dissipation theorem and is based instead on the exact relationship between number fluctuations and density-density correlations. In the low-temperature regime, it allows us to observe significant deviations from fluctuation-dissipation predictions, uncovering sub-extensive fluctuations. Our method is applicable to systems with arbitrary trapping potentials, requiring neither precise trap calibration nor global thermal equilibrium. This nearly universal thermometer for quantum gases overcomes key limitations of existing techniques, paving the way for more accurate and versatile temperature measurements in ultracold quantum systems.\end{abstract}

\maketitle

Ultracold atoms and molecules provide powerful platforms for analog quantum simulation of complex many-body phenomena, where quantum ensembles can be prepared in a variety of energy landscapes, in or out of thermal equilibrium. An essential component of atom- based quantum simulations is the precise measurement of low temperatures, which has been an ongoing fundamental challenge, due the absence of a thermometer that could be universally applied to the diverse many-body problems realized in experiments \cite{kohl2006,zhou2011,roscilde2014,mcdonald2015,mitchison2020}. 

To date, temperature measurements have relied on a range of methods, each with its own strengths and limitations \cite{ketterle1999,gemelke2009,weld2009,sanner2010,muller2010,ku2012,desbuquois2014,mcdonald2015,tobias2020,yan2019,yan2024}. At relatively high temperatures, time-of-flight thermometry is well-suited for systems with a significant thermal fraction, provided ballistic expansion is possible \cite{ketterle1999}. In the ultralow temperature regime, the most common approach involves fitting the in-situ density variations across the trap, assuming the equation of state is sufficiently well known \cite{gemelke2009,hung2011,yefsah2011,yefsah2011a, drewes2016,pasqualetti2024}. When the equation of state is unavailable, fluctuation thermometry as introduced by Zhou and Ho \cite{zhou2011} offers a powerful theory-independent measure of temperature, based on a generalization of the fluctuation-dissipation theorem to trapped systems.  This analysis method is particularly well-suited for experiments with atom-resolved imaging \cite{hartke2020}, but variants of it were used on absorption imaging measurements \cite{muller2010,drewes2016,tobias2020,pasqualetti2024}, including after ballistic expansion \cite{sanner2010}. However, the fitting of density profiles and traditional fluctuation thermometry both face significant limitations: they require precise knowledge of the trapping potential, and they are not applicable to spatially homogeneous systems, essential for studying a number of key quantum phenomena. Additionally, they are unsuitable for measuring local temperatures in out-of-equilibrium scenarios such as thermalization breakdown, thermalization following a quench \cite{gring2012,trotzky2012,cheneau2012,eigen2018}, or heat transport across the system \cite{kim2012,yan2024}.

In this work, we present a novel method for single-atom resolved thermometry of quantum gases, and demonstrate it on an ideal Fermi gas. Using quantum gas microscopy, we precisely measure number fluctuations in a subsystem containing on the order of one atom on average. This enables us to determine both the global and local temperatures across a wide dynamic range. Our approach leverages the exact relationship between number fluctuations and density-density correlations, rather than the fluctuation-dissipation theorem. In the low temperature regime, we observe significant deviations from the fluctuation-dissipation prediction, revealing the presence of sub-extensive fluctuations analogous to those reported in lattice systems \cite{drewes2016,hartke2020,samland2024}. Our method is applicable to any trapping potential, without requiring its precise calibration, and provides a nearly-universal local and global thermometer for quantum gases.

\begin{figure*}[!ht]
    \centering
    \includegraphics[width=0.90\textwidth]{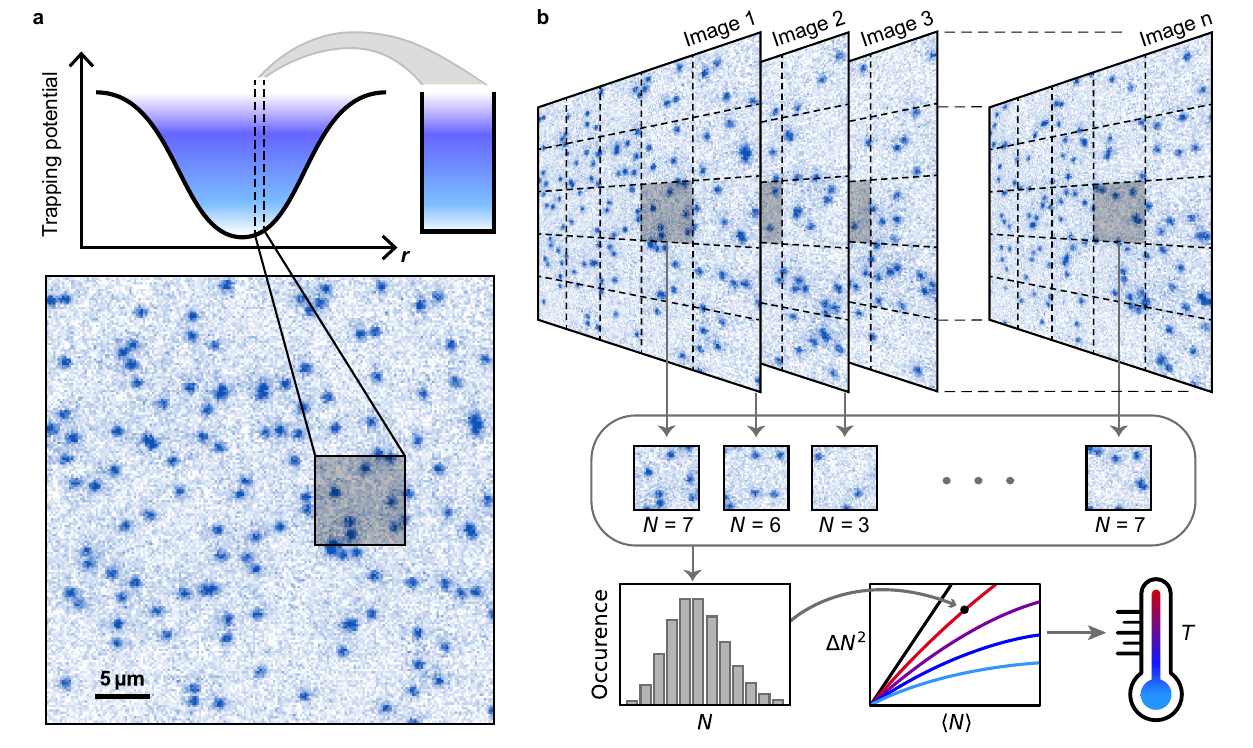}
    \caption{\textbf{Fluctuation Thermometry of an Ideal Fermi Gas.} \textbf{(a)} Bottom: Single-atom image of a two-dimensional, spin-polarized cloud of $^6\mathrm{Li}$ atoms obtained using continuum quantum gas microscopy. Top: The LDA allows taking advantage of the density variations due to trapping potential to analyze the cloud as a collection of homogeneous samples (gray shaded region). \textbf{(b)} Through repeated measurements, we extract the average atom number $\left\langle N \right\rangle$ and variance $\Delta N^2$ for each (quasi-) homogeneous probe volume. For a given $\left\langle N \right\rangle$, the value of $\Delta N^2$ monotonically increases with $T$ such that two joint values $(\left\langle N \right\rangle,\Delta N^2)$ yield a unique temperature. For a thermalized cloud, a full partitioning of the space provides a range of $(\left\langle N \right\rangle,\Delta N^2)$ values all contributing to the same curve, providing accurate thermometry of the entire system.}
    \label{fig:fig1}
\end{figure*}
\section*{Preparing Fermi gases with variable temperature}\label{sec:preparation}

Our experiment starts with a two-component spin (hyperfine state) mixture of $^6$Li atoms confined in a single plane using a `light sheet' dipole trap providing strong confinement in the vertical $z$--direction. In the $xy$-plane, the light sheet provides a shallow nearly harmonic trap, whose spatial variations are slow enough for the local density approximation (LDA) to apply and to extract homogeneous-system quantities \cite{castin2007}. At the ultracold temperatures at which we prepare these systems, collisions only occur through $s$-wave contact interactions, which are forbidden for identical fermions owing to the Pauli-exclusion principle, and can only occur between different spin states. The use of a mixture is therefore required at the beginning of the experiment to ensure cooling and thermal equilibrium. For all preparations, we perform evaporation cooling in the light sheet down to the same trap depth which serves to set the total atom number per spin state to $N_{\rm tot}\approx 150$ and the lowest temperature. We then ramp back up the light sheet to a fixed higher trap depth of $\sim1.8\,\mu$K corresponding to vertical trapping frequency $\omega_z/2\pi\approx3$\,kHz, and heat up the cloud by modulating the light sheet power at a frequency of $\approx2\omega_z$ for a variable time $t_{\rm mod}$. After allowing the system to reach thermal equilibrium for 1.8\,s, we remove one spin component using a spin-selective pushing beam propagating along the vertical direction, ensuring that pushed atoms do not interact with the remaining spin component. The resulting single-component systems realize (quasi-) two-dimensional non-interacting Fermi gases. This protocol allows preparing samples with increasing temperatures but fixed trapping parameters and similar atom numbers. In addition to having eight preparations with $t_{\rm mod}=$[0, 5.5, 12.5, 24, 42, 70, 105, 160]\,ms, we also prepare a classical (hot) gas to serve as a reference. Given the energy spacing $\hbar\omega_z=k_{\rm B}\times143(3)$\,nK in the $z$--direction, we expect population in the excited $z$--levels to increase with temperature. 

To image the clouds, we pin the atoms using an optimized ramp on of a deep optical lattice in the $xy$-plane and apply Raman side-band cooling, which reduces the motional energy of the atoms to the bottom of their respective lattice wells, and simultaneously drives their fluorescence. The photons that are spontaneously emitted by each atom during this process are collected through a high-resolution objective and projected onto a CCD camera. We acquire approximately $1400$ images for each preparation. In Fig.\,\ref{fig:fig1}a, we show a typical experimental image of a non-interacting Fermi gas where each atom is resolved with a near-100\% fidelity. This imaging technique, referred to as continuum quantum gas microscopy, was introduced in our previous work \cite{verstraten2024,jongh2024}.

\section*{Correlation-based Fluctuation thermometry}\label{sec:Fluc_thermo}

\begin{figure*}[!t]
    \centering
    \includegraphics[width=\textwidth]{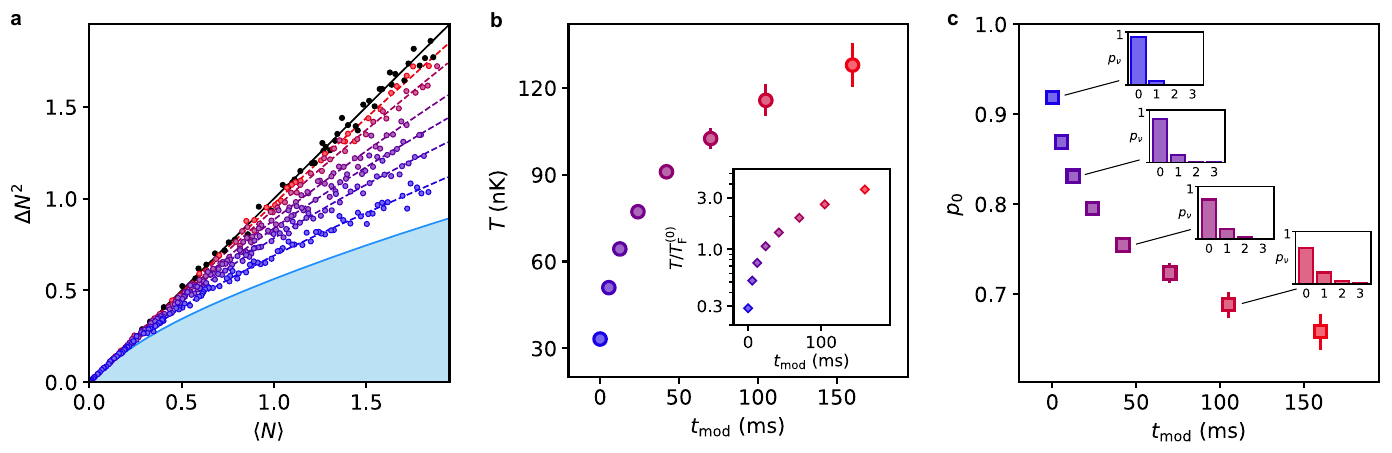}
    \caption{\textbf{Global Thermometry.} \textbf{(a)} 
Variance of the atom number $\Delta N^2$ as a function of the average atom number $\left\langle N \right\rangle$ in each probe volume. Each set of colored points correspond to a different modulation time $t_{\mathrm{mod}}=[0,5.5,12.5,24,70,160]\,\mathrm{ms}$ (from blue to red), and the dashed lines are fits obtained for $T = [33(1), 51(2), 63(1), 78(2), 102(5), 136(14)]\,\mathrm{nK}$, respectively. The set of black points corresponds to a gas prepared in the classical regime, which fall on the expected $\Delta N^2 = \left\langle N \right\rangle$ Poissonian prediction (solid black line) without fitting parameter. The blue solid line corresponds to the theoretical zero temperature fluctuations for an ideal Fermi gas, which are sub-extensive (see corresponding section) and represent the minimal achievable fluctuations. The blue shaded region is physically forbidden for a given average atom number. \textbf{(b)} Cloud temperature $T$ extracted from the fit of the fluctuation data $\Delta N^2 = f(\left\langle N \right\rangle$) for each modulation time $t_{\mathrm{mod}}$. The inset shows the reduced temperatures $T/T_{\mathrm{F}}^{(0)}$  ranging from 0.29(1) to 3.6(3), where $T_{\mathrm{F}}^{(0)}$ is the Fermi temperature at the center of the cloud, demonstrating the applicability of our thermometry technique over a large dynamic range of quantum degeneracy. \textbf{(c)} Fraction of atoms $p_0$ in the transverse vibrational ground state as a function of $t_{\mathrm{mod}}$, and examples of population histograms in the different $z$-levels.}
    \label{fig:fig2}
\end{figure*}

The relation between atom number fluctuations and temperature stems from basic principles. The most common way to establish this connection is via the fluctuation-dissipation theorem for the atom number operator \footnote{The symbol $\hat{}$ is omitted for readability.} $N$ inside a homogeneous probe volume $S$:

\beq
\Delta N^2= k_{\rm B}T\frac{\partial \langle N \rangle}{\partial \mu}\Bigr|_{T}
\label{eq:FDT}
\eeq where $\langle N\rangle$ is average atom number and $\Delta N^2\equiv\langle N^2 \rangle - \langle N \rangle ^2$ is the variance. This result is not exact but sufficiently accurate at \text{non-zero} temperature when the probe volume is large enough (see e.g. \cite{castin2007,astrakharchik2007,klawunn2011}). The link between temperature and number fluctuation can otherwise be established via a more general and exact relation, by recasting the definition of the variance in terms of the density-density correlation: 

\beqa
\Delta N ^2 &=& \iint_{S} d\mathbf{r}_1d\mathbf{r}_2 \,\langle n(r_1)n(r_2)\rangle - \langle N \rangle ^2\nonumber\\  
&=& \langle N \rangle-n^2\iint_{S} d\mathbf{r}_1d\mathbf{r}_2 \,\left[1-g_2(\mathbf{r}_1,\mathbf{r}_2)\right],
\label{eq:varNwithg2}
\eeqa where the two-point reduced density correlation function is:
\begin{equation}
g_2(\mathbf{r}_1,\mathbf{r}_2) =\frac{\langle \psi^{\dagger}(\mathbf{r}_2)\psi^{\dagger}(\mathbf{r}_1)\psi(\mathbf{r}_1)\psi(\mathbf{r}_2)\rangle}{n^2},
\label{eq:g2}
\end{equation} which generically depends on the absolute temperature of the system. The relation is the central point of our approach: if the $g_2$ function is known, Eq.~(\ref{eq:varNwithg2}) provides a thermometer at the length scale of $S$, which can be of any shape or size, provided it is locally at thermal equilibrium. 

For the ideal Fermi gas $g_2$ is fully known \cite{castin2007, jongh2024}, and its functional form directly provides several key insights. First, Eq.~(\ref{eq:varNwithg2}) immediately displays that atom number fluctuations are sub-Poissonian, since $g_2(\mathbf{r}_1,\mathbf{r}_2)<1$ at short distance $r=|\mathbf{r}_1-\mathbf{r}_2|$. Second, since $g_2$ depends only on two dimensionless quantities $k_{\rm F} r$ and $T/T_{\rm F}$ \cite{jongh2024}, with the Fermi wavevector $k_{\rm F}$ and temperature $T_{\rm F}= \hbar^2 k_{\mathrm{F}}^2/(2 m k_{\mathrm{B}})$ only depending on $\langle N \rangle$, the variance $\Delta N ^2$ is effectively a function of $\langle N \rangle$ and $T$. This means that a measurement of two joint values of $\Delta N ^2$ and $\langle N \rangle$ corresponds uniquely to a single value of temperature $T$. In the following we demonstrate the capabilities of our thermometry method to obtain the global and local temperature of ideal Fermi gases, and discuss its broad applicability to quantum gases.

\section*{Global Thermometry}

Dividing our single-atom-resolved images in a square grid gives direct access to local number fluctuations, which we obtain from the many repetitions of a given preparation as depicted in Fig.\,\ref{fig:fig1}b. In the framework of the LDA, probing locally different regions of our non-uniform clouds is equivalent to probing multiple homogeneous systems with varying average atom numbers. Labelling each square $S_{i,j}$ with horizontal and vertical indices $i,j$, we obtain for a given preparation a set of values ($\langle N_{i,j}\rangle$, $\Delta N_{i,j}^2$), yielding a full curve of $\Delta N ^2$ as a function of $\langle N \rangle$. For a thermalized gas, the number fluctuation extracted from each subsystem $S_{i,j}$ should fall on a single curve associated to the temperature $T$. 

In Fig. \ref{fig:fig2}a, we report a set of measurements of $\Delta N^2$ as a function of $\langle N \rangle$, where each data point corresponds to a sub-region $S_{i,j}$ of the cloud. The highest data points (black markers) correspond to the classical gas, which fall on the $\Delta N^2 = \langle N \rangle$ Poissonian line without any fitting parameter. Each set of colored data points correspond to a given modulation time $t_{\rm mod}$, and we see increasingly suppressed density fluctuations with decreasing $t_{\rm mod}$, hence decreasing $T$, as expected. The suppression of density fluctuations for the ideal Fermi gas has been observed in 3D bulk systems via absorption imaging \cite{sanner2010,muller2010,tobias2020} and on a lattice in 2D via quantum gas microscopy \cite{omran2015}. Here, we observe this phenomenon with continuum quantum gas microscopy for the first time, with an unprecedented resolution and precision for a bulk gas. Thanks to the single-atom resolution and near-100\% atom counting fidelity provided by quantum gas microscopy, we obtain direct access to atom number counting and fluctuations, free of the systematic errors inherent to absorption imaging \cite{sanner2010,muller2010,tobias2020}, which require calibrating the absorption cross section and taking into account the photon shot noise of the probe beam. 

In order to extract the temperature we fit each data set of $\Delta N^2$ versus $\langle N \rangle$ using Eq.~(\ref{eq:varNwithg2}), which we adapt to take into account the presence of atoms in higher $z-$levels \footnote{Additional details can be found in the Supplementary Materials, which includes Ref. \cite{jin2024}.}, with $T$ as the sole fitting parameter. For each preparation, all sub-regions $S_{i,j}$ of the cloud contribute to the fitted data, making the method particularly robust to obtain the global temperature and sensitive to deviations to thermalization if present in the system. The resulting values of $T$, the reduced temperature at the center of the trap $T/T_{\rm F}^{(0)}$, and population $p_0$ of the $z-$level ground state, are reported in Fig.\,\ref{fig:fig2}b-c.

As a check of reliability for the obtained temperatures, we compute the theoretical $g_2$-function expected for each previously fitted temperature and compare the result to the $g_2$-function measured from the same samples, obtaining excellent agreement without any fitting parameter \cite{Note2}.  

\section*{Local Thermometry}

\begin{figure}[!t]
    \centering
    \includegraphics[width=\columnwidth]{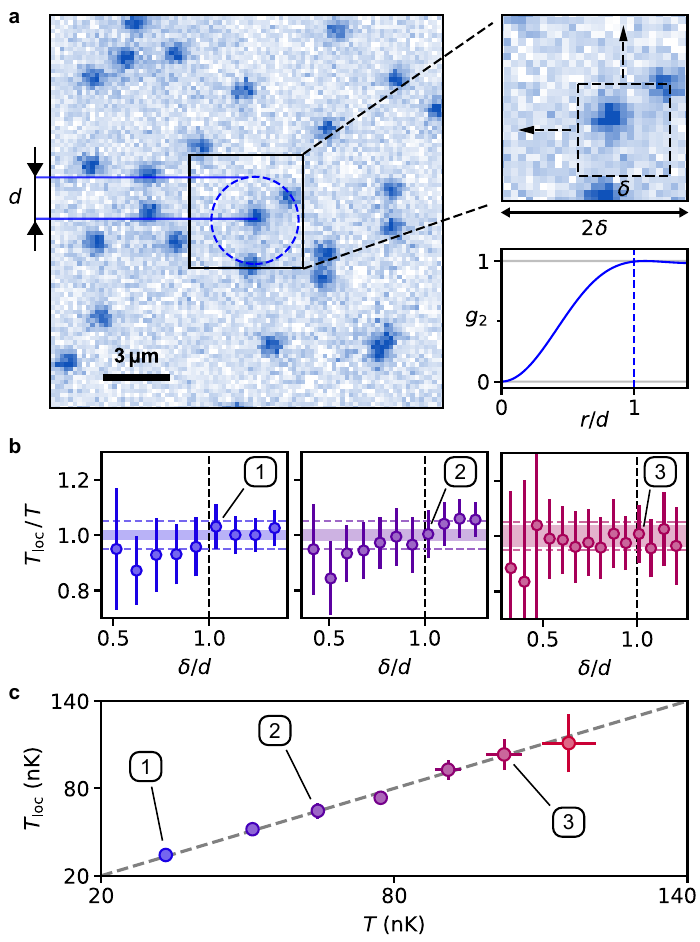}
    \caption{\textbf{Local Thermometry.} \textbf{(a)} Atom number fluctuations are analyzed within a small square-shaped probe volume near the center of the cloud. To obtain more than a single a pair of values ($\langle N \rangle, \Delta N^2$), we introduce a smaller square volume of size $\delta$ inside the probe volume (top right panel). Scanning its position yields an ensemble of temperature measurements. To reveal the local nature of the extracted temperature, the probe volume size $2\delta$ is varied over a range on the order of the Fermi hole diameter $2d$ at $T=0$. As an illustration, the Fermi hole is shown as a blue dashed circle centered on one atom inside the probe volume. The disk represents the correlations region of that atom, as can be seen from the functional form of the second order correlation function $g_2$ (bottom right panel). \textbf{(b)} Extracted local temperature $T_{\rm loc}$ as a function of the ratio $\delta/d$ for three of our samples. The temperatures are normalized to the global temperature $T$ obtained from the whole sample. The colored band reprents the uncertainty on $T$ and the horizontal dashed lines the $\pm$5\% distance from the mean. The vertical black dashed lines indicate the point where the probe volume size is equal to the correlated region diameter. \textbf{(c)} $T_{\rm loc}$ as a function of $T$ for $\delta\approx d$, showing excellent agreement over a large dynamic range.}
    \label{fig:Tloc}
\end{figure}

According to Eq.~(\ref{eq:varNwithg2}), there is no formal limitation to how small the system $S$ can be, and one can in principle obtain the local temperature at the scale of the correlation length. For an ideal Fermi gas, the correlation length is given by the radius of the Fermi hole, which corresponds well to the inter-particle spacing $d=1/\sqrt{n}=\sqrt{4\pi}k_{\rm F}^{-1}$, with $n$ the density of the gas, near $T=0$ (see bottom right panel of Fig.\,\ref{fig:Tloc}a). To demonstrate this, we apply our method on a single probe volume near the center of the cloud, of variable size $2\delta$ on the order of $2d$ as illustrated in Fig.\,\ref{fig:Tloc}a. We here use a slightly altered version of the analysis performed in the previous section, where a single probe volume provided a single point in the $\langle N \rangle, \Delta N^2$ plane. We introduce a smaller square box of size $\delta$ within the probe volume and scan its position, as depicted in Fig.\,\ref{fig:Tloc}a (top right panel). For each position of the small square, we extract $\langle N \rangle, \Delta N^2$ and hence a temperature, which provides us with a collection of temperature measurements, from which we obtain a confidence interval \cite{Note2}. 

In Fig.\,\ref{fig:Tloc}b, we show typical results for the ratio $T_{\rm loc}/T$ of the measured local temperature to the previously extracted global temperature, as a function of $\delta/d$. This demonstrates that our approach allows accurate local thermometry, down to a fraction of $d$. In Fig.\,\ref{fig:Tloc}c, we summarize the result of our analysis by comparing the values of $T_{\rm loc}$ obtained for $\delta\approx d$ to $T$ for different samples covering a large dynamic range ($T/T_{\rm F}^{(0)}=0.29(1)$ to 2.6(2)), finding excellent agreement.   

In most previous works on fluctuation-thermometry \cite{muller2010,sanner2010,tobias2020,pasqualetti2024}, probing volumes at the scale of the correlation length $d$ would have not been possible due to limited imaging resolution and signal-to-noise. In Ref. \cite{hartke2020}, quantum gas microscopy of Fermi-Hubbard gases allowed to perform thermometry at the scale of several times the correlation length, using the generalized fluctuation-dissipation theorem \cite{zhou2011}.

\section*{Sub-extensive fluctuation measurement}

We now leverage our thermometry method to reveal the presence of sub-extensive number fluctuations in our samples and quantitatively study their behavior as a function of temperature. For a quantum system, the exact relation that relates $\Delta N^2$ to the compressibility inside a homogeneous probe volume $S$ reads \cite{Note2}:
\beq
\Delta N^2= k_{\rm B}T\frac{\partial \langle N \rangle}{\partial \mu}\Bigr|_{T}+\Delta_Q.
\label{eq:FDTdq}
\eeq The correction term $\Delta_Q$ is positive and captures sub-extensive atom number fluctuations, i.e., scaling slower than $N$. In the thermodynamic limit and at finite temperature $\Delta_Q$ is negligible, which yields Eq.\,(\ref{eq:FDT}). When the size of the probe volume is small, as is required for local thermometry, sub-extensive number fluctuations are an important contribution. At $T=0$, $\Delta_Q$ is the only contribution to the number fluctuation, and coincides with the \textit{quantum variance} introduced by Fr{\'e}rot and Roscilde \cite{frerot2016}. For free fermions in $d$--dimensions, $\Delta_Q(T=0)$ scales as $L^{d-1}\log(L^d)$ \cite{gioev2006,castin2007,astrakharchik2007,smith2021}, reflecting the violation of the area law for the entanglement entropy \cite{gioev2006,wolf2006,song2012,calabrese2012,frerot2015}. At finite temperature $\Delta_Q$ also contributes but its behavior is poorly understood beyond the study of specific cases \cite{klawunn2011,frerot2016}, though it is generally expected to vanish as the temperature increases. The correction $\Delta_Q$ corresponds to the non-local atom number correlations observed in Fermi-Hubbard gases \cite{drewes2016,hartke2020,samland2024}. 

\begin{figure}
    \includegraphics[width=0.48\textwidth]{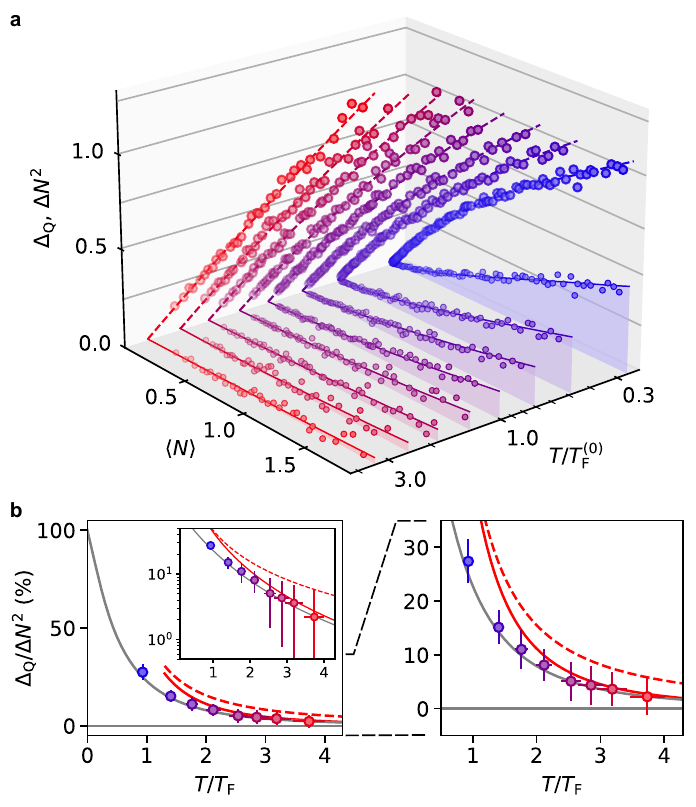}
    \caption{\textbf{Sub-extensive atom number fluctuations.} \textbf{(a)} Atom number fluctuations $\Delta N^2$ (large circles) and sub-extensive contribution $\Delta_Q$ (small circles). Each color corresponds to a different modulation time ranging from $t_{\mathrm{mod}} = 0\,\mathrm{ms}$ (blue) to $t_{\mathrm{mod}} = 160\,\mathrm{ms}$ (red). The $\Delta_Q$ data is obtained by subtracting  the extensive term $k_{\mathrm{B}}T \frac{\partial \left\langle N \right\rangle}{\partial \mu}|_{T}$ from the total fluctuation in Eq.\,(\ref{eq:FDTdq}). The dashed lines going through to $\Delta N^2 = f(\left\langle N \right\rangle$) data are fits as already shown in Fig. \ref{fig:fig2}, whereas the solid lines going through the $\Delta_Q = f(\left\langle N \right\rangle$) data is the theoretical prediction from Eq. (\ref{eq:dq}) without fitting parameter. \textbf{(b)} Relative sub-extensive contribution $\Delta_Q$ for a given probe volume of fixed size and average atom number as a function of the reduced temperature $T/T_{\mathrm{F}}$, shown in linear (main left panel) and semi-logarithmic scale (inset), alongside a zoomed-in version (right panel). The solid and dashed red lines correspond to the high temperature expansion ($T/T_{\mathrm{F}} \gg 1$) with excited $z$-levels populations and in the purely 2D case (Eq. \ref{eq:high_T_deltaQ}). The grey lines are obtained from an exact numerical computation using Eq.\,(\ref{eq:FDTdq}) and including the known population of the excited $z$-levels.}
    \label{fig:Dq}
\end{figure}

Here, we use our correlation-based fluctuation-thermometry, to disentangle the fluctuation-dissipation theorem and $\Delta_Q$ contributions. Having extracted the density and the temperature of our samples, we compute $\partial \langle N \rangle/\partial \mu\bigr|_{T}$ locally \cite{Note2}, and use Eq.~(\ref{eq:FDTdq}) to obtain the value of $\Delta_Q$ in each probe volume $S_{i,j}$ of the cloud. In Fig. \ref{fig:Dq}a we report the measurement of $\Delta_Q$ as a function of $\langle N \rangle$ for the different preparations discussed above. The results confirm that the sub-extensive fluctuations constitute an important fraction of the variance at low temperature. To quantify the temperature dependence of $\Delta_Q$ we select, from the different preparations, probe volumes of fixed square shape and size ($L=2.2\,\mu$m $=2.1(2)\,k_{\rm F}^{-1}$), with similar average densities. The reduced temperature $T/T_{\rm F}$ is therefore the only varying parameter. Results are shown in Fig.\,\ref{fig:Dq}b, where we observe that $\Delta_Q/\Delta N^2$ monotonically decreases to zero with increasing temperature, which is the expected qualitative behavior \cite{klawunn2011,frerot2016}. This shows that even for a small system, the number fluctuations converge to the result of the fluctuation-dissipation theorem at high temperature. Similar behavior was observed in Fermi-Hubbard gases \cite{drewes2016}.

To gain deeper insight on sub-extensive number fluctuations, we derive the following general expression \cite{Note2} considering a total system $\Omega=S+\bar{S}$ that is well described by the grand-canonical ensemble: 

\beq
\Delta_Q =n^2\int_{S} d\mathbf{r}_1\int_{\bar{S}} d\mathbf{r}_2 \,\left[1-g_2(\mathbf{r}_1,\mathbf{r}_2)\right].
\label{eq:dq}
\eeq This relation is valid as long as the size of $\bar{S}$ exceeds the one of the probe volume $S$ by several times the correlation length, which immediately signals that $\Delta_Q$ is a non-local quantity that depends on the shape and size of $S$. It also shows that the zero- and finite-temperature contributions to sub-extensive fluctuations have the same microscopic origin, as they result from the geometric interplay between the boundaries of the probe volume and the functional form of the $g_2$ correlation function. For the (quasi-)2D ideal Fermi gases considered here $\Delta_Q$ can be computed at any temperature by numerically integrating Eq.\,(\ref{eq:dq}) \cite{Note2}. We compare the obtained result to the data in Fig.\,\ref{fig:Dq}b showing excellent agreement. By using the expression of $g_2$ at high temperature \cite{Note2}, we also obtain an analytical expression for the 2D Fermi gas for $T/T_{\rm F}\gg1$ and square probe volume of size $L\gg\lambda_{\rm T}$:

\beq
\Delta_Q\sim\left(\frac{T_{\rm F}}{T}\right)^{3/2}\sqrt{\frac{N}{2\pi^2}},
\label{eq:high_T_deltaQ}
\eeq showing that the sub-extensive character survives at high temperature, but with a suppressed relative weight compared to the fluctuation-dissipation contribution. We generalize this prediction to take into account non-zero populations in higher $z$-levels \cite{Note2}. The result is compared to our measurements in Fig.\,\ref{fig:Dq}b, showing excellent agreement with the observed asymptotic behavior.

\section*{Discussion}

Our atom-resolved thermometry method only relies on the exact relation between the number fluctuation and the density-density correlation function $g_2$, and as such provides both a global as well as a local thermometer for a wide range of quantum systems. It universally applies to any system provided the temperature dependence of the $g_2$-function can be computed. This holds not only for ideal quantum gases where analytical results exist, but also quite broadly for interacting 1D systems with contact \cite{zabrodin1985,berkovich1987,leclair1996,leclair1999,kheruntsyan2003,caux2006} and long-range interactions \cite{dzyaloshinskii1996}, 2D BKT superfluids \cite{ceperley1995,pilati2005}, 3D Bose gases \cite{holzmann1999}, Bose-Hubbard systems with short-range \cite{capogrosso-sansone2008,caleffi2020} and long-range interactions \cite{batrouni2000,capogrosso-sansone2010,yamamoto2012}, as well as attractive and repulsive Fermi-Hubbard systems \cite{zhang1999,varney2009,paiva2010,cheuk2016,he2019,chan2020}. More generally, the density-density correlation function can in principle be computed numerically for any interacting system for which unbiased QMC methods exists. This includes most bosonic systems, and a variety of fermionic systems on lattices and in the continuum (see e.g \cite{pollet2012,chang2015,alexandru2022,carlson2011,he2019,vanhoucke2012,rossi2018}). 

Our correlation-based thermometry also readily applies to systems connected to a `trivial' reservoir. This situation is commonly realized in spin-imbalanced 2D and 3D BEC-BCS crossover superfluids, where the system phase-separates, with an ideal Fermi gas forming on the outer part of the trap \cite{Shin2006,Zwierlein2006,Shin2008,Mitra2016}. In quantum simulation of the Fermi-Hubbard model, reservoir-based quantum state engineering is crucial to achieve low temperatures \cite{mazurenko2017, chiu2018}. In all these cases, one can study the strongly-correlated system near the center of the trap, and use the trivial-state reservoir to perform thermometry as demonstrated here. 

\section*{Conclusion}\label{sec:conclusion}

In this work, we demonstrated a novel method for single-atom resolved thermometry of quantum gases using quantum gas microscopy. By leveraging the exact relationship between number fluctuations and density-density correlations, our approach overcomes key limitations of traditional thermometry, including the dependence on precise trap calibration and global equilibrium. This enabled us to accurately measure both global and local temperatures across a broad dynamic range and uncover sub-extensive fluctuations in low-temperature regimes. Our nearly universal method complements fluctuation thermometry based on the fluctuation-dissipation theorem \cite{zhou2011,hartke2020} and extends thermometry to quantum gases far out-of-equilibrium \cite{gring2012,trotzky2012,cheneau2012,eigen2018, schreiber2015,choi2016,alet2018,abanin2019,kim2012,yan2024}. Finally, our work highlights the importance and need for further theoretical studies to accurately compute spatial density-density correlations in many-body systems.

\section*{Acknowledgements}

We thank Yvan Castin, Ir\'en\'ee Fr\'erot, Fabrice Gerbier, and F\'elix Werner for insightful discussions, and Ir\'en\'ee Fr\'erot for
a critical reading of the manuscript. We are grateful to Antoine Heidmann for his crucial support as head of Laboratoire Kastler Brossel. This work has been supported by Agence Nationale de la Recherche (Grant No. ANR-21-CE30-0021), CNRS (Tremplin@INP 2020), and R{\'e}gion Ile-de-France in the framework of DIM SIRTEQ and DIM QuanTiP. 

\textit{Note} -- While writing the manuscript, we became aware of other work on quantum gas microscopy in the continuum \cite{yao2024,xiang2024}.

\newpage

\renewcommand\thefigure{S\arabic{figure}}
\renewcommand\theHfigure{S\arabic{figure}}
\setcounter{figure}{0} 

\renewcommand\theequation{S\arabic{equation}}
\renewcommand\theHequation{S\arabic{equation}}
\setcounter{equation}{0} 

\noindent

\section*{Supplementary Materials}\label{sec:Methods}

\subsection*{Experimental Sequence}
Our experimental setup was described in \cite{verstraten2024}. For this work, we start by preparing a spin-balanced mixture of $^6$Li atoms composed of the two lowest-energy hyperfine states, denoted $\ket{1}$ and $\ket{2}$. The gas is loaded into a light sheet potential described in the main text, where we perform evaporative cooling at 832\,G, corresponding to the Feshbach resonance of the mixture. The magnetic field is then reduced to $585$\,G where we transfer the state $\ket{2}$ atoms to the third-lowest hyperfine state $\ket3$, via a radio-frequency Landau-Zener sweep. At this magnetic field, the $\ket{1}-\ket{2}$ and $\ket{1}-\ket{3}$ scattering lengths are equal, preventing any density-related effect on the transition. The magnetic field is then ramped to $320$\,G where $a_{13} = -950\,a_0$, with $a_{13}$ the $\ket{1}-\ket{3}$ scattering length and $a_0$ the Bohr radius. We continue the evaporation by lowering the light sheet depth down to $\sim$40\,nK for a duration of $\sim 14$\,s. Atoms are held in the light sheet for 1.6\,s before adiabatically ramping the trap depth back to $\sim 650$\,nK in $300$\,ms, corresponding to $\omega_z = 2\pi \times 3.0(1)$\,kHz. Then the light sheet intensity is modulated for a variable time $t_{\rm mod}$, with a frequency $\omega_{\rm mod} \simeq 2\,\omega_z$ and an amplitude of 5\,\% the trap depth. Following this controlled heating, we hold the atoms in the trap for 1.8\,s for thermalization. Afterwards the atoms in the state $\ket{1}$ are expelled using resonant light propagating directed downwards. Because of the dilution and the weak interactions, the removal does not affect the atoms in state $\ket{3}$. Finally the magnetic field is ramped to 0\,G in $15$\,ms and the remaining atoms are pinned according to the protocol described in \cite{verstraten2024,jongh2024}.\\

The pinning lattice is formed by the self-interference of a red-detuned 1064\,nm laser, crossing three laser arms at 120$^\circ$ angles in the horizontal plane and creating a triangular lattice with a spacing of 709\,nm \cite{jin2024}. At the highest laser power, the trapping frequency of the lattice wells is measured to be 1.0(1)\,MHz. Upon pinning, the power of both the light sheet and the $xy$-lattice are increased to their maximum power in $\sim10\,\mu$s, immediately followed by Raman sideband cooling (RSC)\cite{verstraten2024}. The fluorescence signal is captured by exposing a camera after a 2\,s hold period to minimize fluctuations in background light. The resulting images are processed using a highly accurate neural network recognition algorithm to determine the positions of individual atoms \cite{verstraten2024}. For each experimental run, two images are captured with an exposure time of 600\,ms and a 250\,ms interval between exposures, all while keeping the RSC and lattice beams active. This setup enables real-time measurement of the fraction of atoms that remain pinned in their lattice sites, achieving a rate exceeding 99.9\,\%.

\subsection*{Thermalization} 

Following the light sheet power modulation, we let the atoms thermalize for a duration of 1.8\,s. We quantitatively determine this thermalization time by preparing a mixture of $\ket{1}$ and $\ket{3}$ atoms with the same evaporation procedure as presented previously and modulating the light sheet trap power for a 160\,ms duration, the longest modulation time used for our samples, corresponding to the least favorable thermalization condition. As the modulation is only resonant with respect to the vertical trapping frequency, no energy would be injected along the horizontal plane in the absence of interpaticle interactions.

\begin{figure}[!h]
    \centering
    \includegraphics[width=0.4\textwidth]{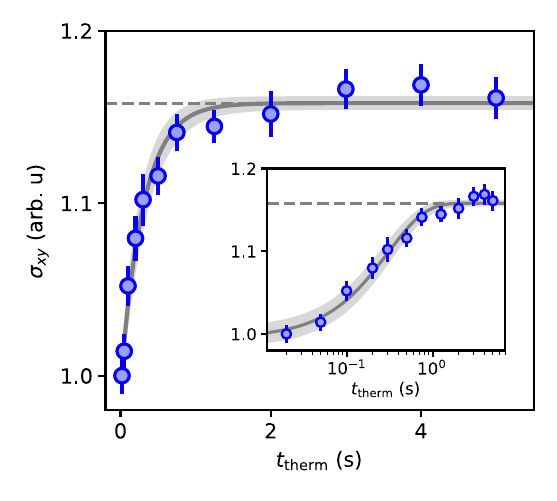}
	\caption{\textbf{Thermalization Curve.} In-plane cloud width ($\sigma_{xy}$) of atoms in the light-sheet as a function of thermalization time $t_{\rm therm}$, directly after a 160 ms power modulation of the trap (data points). The error bars correspond to the standard error of the distribution. The gray line is an exponential fit to the data which reaches 99\,\% of its asymptotic value after 1.4\,s. With the shaded gray area we present the uncertainty of the fit. The inset shows the same figure with a logarithmic $x$-axis.}
    \label{fig:thermalization}
\end{figure}

We vary the thermalization time $t_{\rm therm}$ during which we hold the gas after the trap depth modulation. Then we remove the state $\ket{1}$ atoms, turn on the pinning lattice, perform RSC and capture a single atom image. In Fig.\,\ref{fig:thermalization} we present the influence of $t_{\rm therm}$ on $\sigma_{xy} = \sqrt{\sigma_{x}^2 + \sigma_{y}^2}$, with $\sigma_{x}$ (resp. $\sigma_{y}$) the standard deviation of the single-atom distribution along the horizontal axis $x$ (resp. $y$). The figure shows that the collisions between state $\ket{1}$ and state $\ket{3}$ atoms lead to a redistribution of the energy from the vertical direction to the horizontal plane with a relaxation time ($\sim1.4$\,s) smaller than our holding time (1.8\,s), ensuring that our samples are well thermalized. The observation that the number fluctuation measurements $\Delta N^2$ fall on the same temperature curve for the full range of $\langle N \rangle$ provides an independent additional indication that we have reached thermal equilibrium after modulation.

\begin{figure*}[!t]
    \centering
    \includegraphics[width=\textwidth]{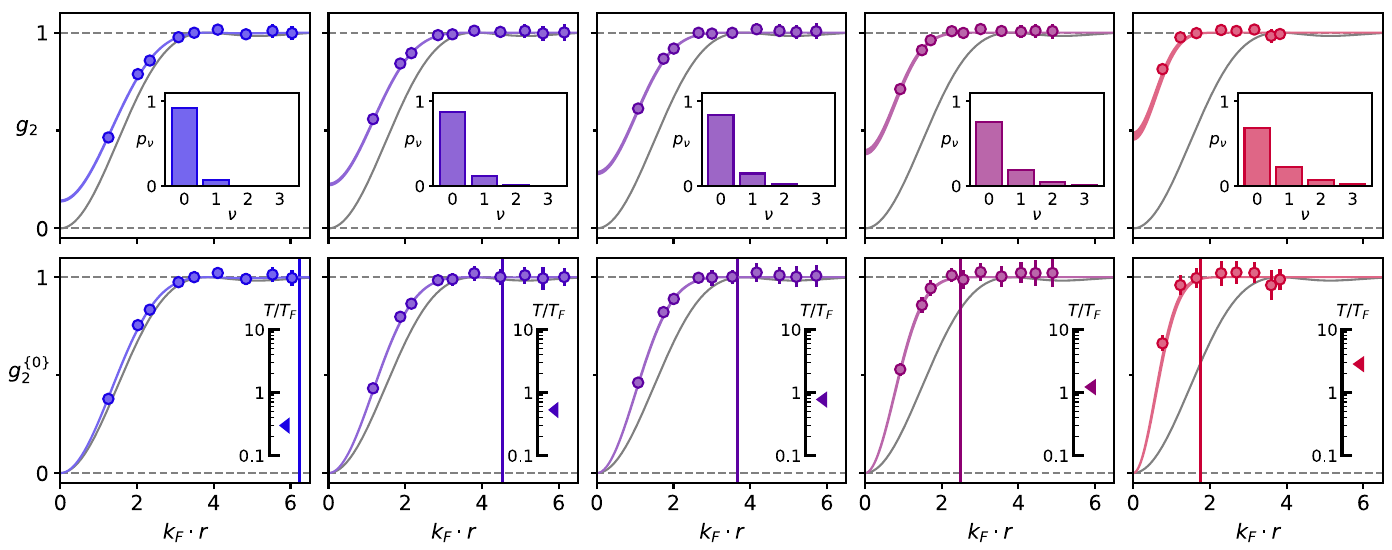}
	\caption{\textbf{$g_2$ correlation functions.} First row: normalized two-point correlation function $g_2$ as described in \cite{jongh2024} for the modulation times $t_{\rm mod} = [0,5.5,12.5,42,105]$\,ms in the central region of the trap. With the shaded colored regions we include the theoretical correlation functions at the temperatures extracted from the fluctuation thermometry, with the estimated temperature error included in the thickness of the curve. The gray solid lines represent the zero temperature limit. The insets show histograms of the vertical-levels populations inside the central region. Second row: estimation of the $g_2$ functions for the atoms in the vertical ground state, with the theoretical prediction shown by shaded colored regions. The vertical lines indicate the thermal de Broglie wavelength $\lambda_{\rm T} = h/\sqrt{2\pi m k_{\rm B} T}$ calculated with the estimated temperatures, highlighting that, as expected, the correlation length converges towards $\lambda_{\rm T}$ at high $T/T_{\rm F}$. As insets for the second row figures, we point at the $T/T_{\rm F}$ value obtained from global density fluctuation thermometry inside the region used to extract the correlation function.}
    \label{fig:g2}
\end{figure*}

\subsection*{Thermometry Analysis}

While the LDA can be applied in the horizontal plane  of the light sheet potential, the large vertical trapping frequency requires a fully quantized treatment of the motion in that direction. The occupation in the motional $z$-levels of quantum number $\nu$ is given by the expression:

\beq
p_\nu = \frac{T}{T_{\rm F}} \ln{\left(1+e^{(\mu - \nu \hbar\omega_z)/k_{\rm B} T}\right)},
\label{eq:p_nu}
\eeq with $\mu$ the chemical potential. For a given vertical frequency $\omega_z$ (which we experimentally determine via parametric heating measurement), the distribution among the vertical harmonic oscillator levels is solely dependent on density and temperature.

Using Eq.~(\ref{eq:varNwithg2}) we assign a measurement of both $\Delta N$ and $\langle N \rangle$ to a single value of the temperature $T$ in the two dimensional case. In other words, the number fluctuations of a purely 2D (i.e., for which  only a single vertical level is occupied) non-interacting Fermi gas are given by a numerically computable function: $\Delta N^2 = f\big(\langle N \rangle, T\big)$. In the case where many vertical $z-$levels are populated and writing $N_\nu$ the particle number in the state $\nu$, the atom number variance of the gas results from the sum of the atom number variance within each $z-$level:

\beqa
\Delta N^2 &=& \sum_\nu \Delta N_\nu^2\nonumber\\
&=& \sum_\nu f\big(\langle N_\nu \rangle, T\big)\nonumber\\
&=&\sum_\nu f\big(p_\nu\langle N \rangle, T\big).
\label{eq:varNQuasi2D}
\eeqa As explained above, $p_\nu$ only depends on the temperature and the average density (proportional to $\langle N \rangle$). Consequently, $T$ is still the only free parameter in the fitting of the $\Delta N^2$ versus $\langle N \rangle$ points.

\subsection*{Comparison with the Correlation functions}

In order to determine the reliability of the obtained temperatures, we compute the theoretical $g_2$-function expected for each temperature determined from our number fluctuation measurement and compare the result to the $g_2$-function extracted from the same samples without any fitting parameter, taking into account the vertical states according to the method described in \cite{jongh2024}. The result is shown in the first row of Fig.\,\ref{fig:g2}, displaying an excellent agreement throughout the full dynamic range of temperatures probed. The observed non-zero values of $g_2(0)$ are a result of the higher occupation of the excited $z-$levels, which increases with $T$. As was done previously \cite{jongh2024}, we also extract the $g_2$-function of the $\nu=0$ state $g_2^{\lbrace 0\rbrace}(r)$ as reported on the second row of Fig.\ref{fig:g2}.

\subsection*{Influence of the Probe Volume on the Global Thermometry}

\begin{figure*}[!t]
    \centering
    \includegraphics[width=\textwidth]{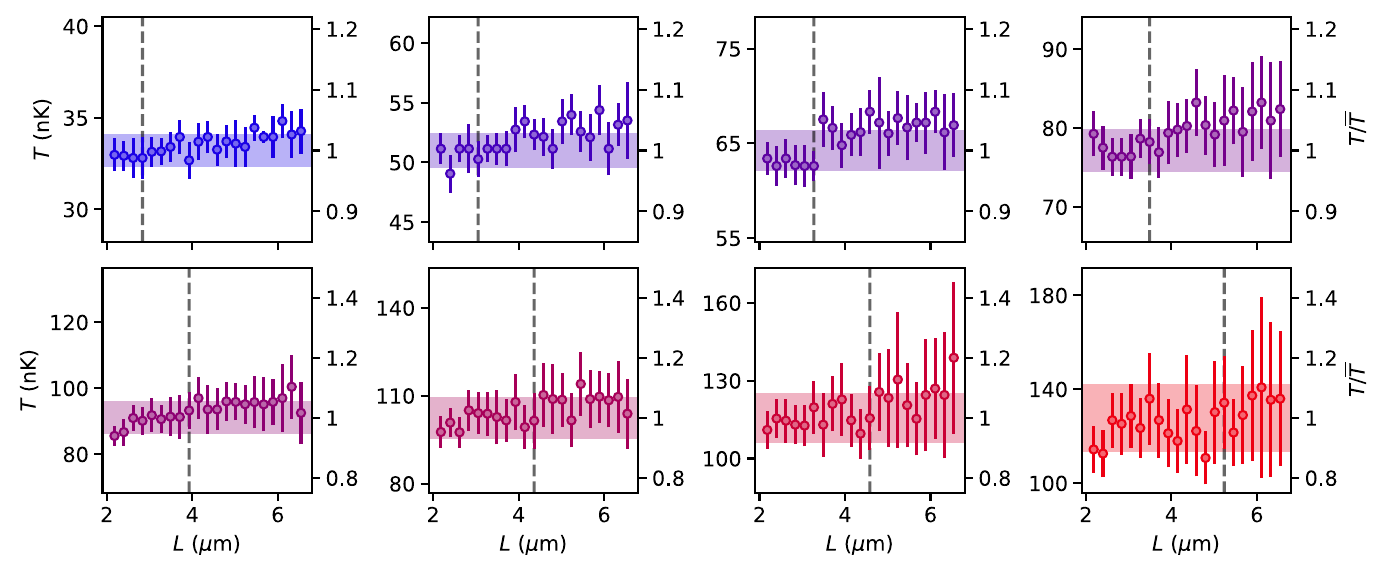}
	\caption{\textbf{Probe volume influence on the global temperature estimation.} Global temperature estimation for varying square probe sizes $L$. Each panel correspond to a different modulation time ranging from $t_{\mathrm{mod}}=0$\,ms (top left panel) to $t_{\mathrm{mod}}=160$\,ms (bottom right panel). The filled area of each figure is our confidence interval, that we define as the average of the confidence intervals obtained with probe volumes of sizes ranging from 2.6\,$\mu$m to 3.7\,$\mu$m. The vertical dotted lines are positioned at the $L$ values used in Fig.\,\ref{fig:fig2}a.}
    \label{fig:VaryBinSize}
\end{figure*}

For a homogeneous system, there are no fundamental constraints on the size $L$ of the square probe volumes used to partition our images when estimating the temperature via the density fluctuations. In practice, however, we identified two constraints. Firstly, there is a lower limit on $L$ due to the pinning lattice, essential for the imaging process. Using a probe volume $L\lesssim a_L$ would bias the atom number distribution as some sub-regions would have more lattice sites than others. We have observed that this effect is marginal for $L\gtrsim 3\,a_L\approx 2\,\mu$m.
Secondly, an upper bound on probe volume size stems from trap inhomogeneities. The presence of a density gradient inside the probe volume, due to a spacial variation of the potential, will lead to an increase of the density variance, adding to the intrinsic fluctuation. This will cause an overestimation of the local variance with respect to the average density, and thus to systematically overestimate the temperature.
In Fig.\,\ref{fig:VaryBinSize} we show the estimated temperatures as a function of the square probe size $L$, with errors intervals obtained via bootstrapping. These results show that, over a broad range of probe sizes (2$\mu$m $\lesssim L \lesssim 6\mu$m), the measured temperature remains within $\pm5$\,\% of the temperature obtained in the safe range, which we determined to be 2.6$\mu$m $\leq L \leq 3.7\mu$m.

\subsection*{Local Temperature Error Estimation}

In Fig.\,\ref{fig:Tloc} we measure the local temperature by scanning the position of a small square of size $\delta$ across all allowed positions, with one-pixel steps, within the larger probe volume of size $2\delta$. At each position $i$ of the smaller square, we evaluate the temperature $T_i$ using our fluctuation thermometry method. $T_{\rm loc}$ is determined by taking the median value of the $(T_i)$ distribution. From this distribution of $(T_i)$ values we extract the standard deviation $\Delta T$.
With this procedure, the temperature estimates are not independent due to the correlations among the particle numbers of the various probes. Within the $2\delta$-sized square probe volume, there are four non-overlapping smaller squares. We therefore divide $\Delta T$ by $\sqrt{4}$ to estimate the error on the $T_{\rm loc}$ values, presented in Fig.\,\ref{fig:Tloc}. In the case of a classical gas, where there is no correlation between the particle number inside and outside a given volume, non-overlapping regions have independent number fluctuations and the division is pertinent. For a fermionic gas, however, the atom numbers in two adjacent boxes are anti-correlated, leading to anti-correlations between the corresponding temperature measurements, reducing the standard error. The error bars given above are therefore conservative estimates.

\subsection*{Sub-Extensive Fluctuation}

In the grand canonical ensemble, one can derive:

\beq
\Delta N_S^2 = k_{\rm B}T \frac{\partial\langle N_S \rangle}{\partial \mu}\Bigr|_{T} - \textrm{cov}(N_S,N_{\overline{S}}).
\label{eq:VarN_equal_Comp_minus_Cov}
\eeq with $N_S$ the atom number in the volume $S$, $N_{\overline{S}}$ the atom number outside $S$, $\mu$ the chemical potential. $\Delta_Q =  - \textrm{cov}(N_S,N_{\overline{S}})$ is the correction to the fluctuation-dissipation theorem (Eq.~(\ref{eq:FDT}) of the main text). To extract this sub-extensive contribution from our data, we use the measured temperature to determine the chemical potential via Eq.~(\ref{eq:p_nu}). Then using the expression of $\langle N_S \rangle$ in a quasi-2D homogeneous system:

\beq
\langle N_S \rangle = \frac{S}{\lambda_{T}^2} \sum_\nu \log{\left(1+e^{(\mu - \nu \hbar\omega_z)/k_{\rm B} T}\right)},
\label{eq:Mean_NS}
\eeq we compute:

\beq
k_{\rm B} T \frac{\partial\langle N_S \rangle}{\partial \mu}\Bigr|_{T} = \frac{S}{\lambda_{T}^2} \sum_\nu \frac{1}{1+e^{-(\mu - \nu \hbar\omega_z)/k_{\rm B} T}}.
\label{eq:dN_over_dmu_with_sum}
\eeq Knowing the temperature, we can numerically calculate the right hand side and use Eq.~(\ref{eq:VarN_equal_Comp_minus_Cov}) to extract the values of $\Delta_Q$ of Fig.\,\ref{fig:Dq}.\\

To derive Eq.~(\ref{eq:dq}) in the main text, we start from our definition of $\Delta_Q$:

\beqa
\Delta_Q &=& - \textrm{cov}(N_S,N_{\overline{S}})\nonumber\\
&=& \langle N_S \rangle\langle N_{\overline{S}} \rangle - \langle N_S\,N_{\overline{S}} \rangle\nonumber\\
&=& \langle N_S \rangle\langle N_{\overline{S}} \rangle - \int_{S} d\mathbf{r}_1\int_{\bar{S}} d\mathbf{r}_2 \,\langle n(\mathbf{r}_1) n(\mathbf{r}_2)\rangle\nonumber\\
&=& n^2 \int_{S} d\mathbf{r}_1\int_{\bar{S}} d\mathbf{r}_2 \,\left[1-g_2(\mathbf{r}_1,\mathbf{r}_2)\right].
\label{eq:dq_derivation}
\eeqa

This expression demonstrates that $\Delta_Q>0$ for polarized fermions ($g_2<1$). In the context of a non-interacting Fermi gas, Wick's theorem connects $g_2$ to the first order correlation function $g_1(\mathbf{r}_1,\mathbf{r}_2) = \langle \psi^{\dagger}(\mathbf{r}_1) \psi(\mathbf{r}_2)\rangle/n$ and gives: $g_2 = 1 - g_1^2$. Using that relation we obtain:

\beq
\Delta_Q = n^2 \int_{S} d\mathbf{r}_1\int_{\bar{S}} d\mathbf{r}_2 \,g_1(\mathbf{r}_1,\mathbf{r}_2)^2.
\label{eq:dq_int_g1}
\eeq

We used the expression above to numerically integrate $\Delta_Q$ in Fig.\,\ref{fig:Dq}. For $T/T_{\rm F}\rightarrow \infty$, one can show that: $g_1(k_{\rm F} r) \simeq e^{-\pi r^2/\lambda_T^2}$. If then we consider for $S$ a square volume of size $L \gg \lambda_T$, we get for the purely 2D case:

\beq
\Delta_Q^{2D} \simeq \left(\frac{T_{\rm F}}{T}\right)^{3/2} \sqrt{\frac{N}{2\pi^2}}.\\
\label{eq:highT_dq_derivation}
\eeq

In the case where the vertical states are populated:

\beqa
\Delta_Q &= & \sum_\nu p_\nu^2 \Delta^{2D}_{Q, \nu} \nonumber\\
&\simeq & \sum_\nu p_\nu^2 \left(\frac{T_{\rm F} p_\nu}{T}\right)^{3/2}\sqrt{\frac{N p_\nu}{2\pi^2}}\nonumber\\
&\simeq & \left( \sum_i p_\nu^4 \right) \left(\frac{T_{\rm F}}{T}\right)^{3/2}\sqrt{\frac{N}{2\pi^2}}
\label{eq:highT_quasi2D_dq_derivation}
\eeqa

The expression above is shown with a red solid line in Fig.\,\ref{fig:Dq}b.


\begin{thebibliography}{82}%
\makeatletter
\providecommand \@ifxundefined [1]{%
 \@ifx{#1\undefined}
}%
\providecommand \@ifnum [1]{%
 \ifnum #1\expandafter \@firstoftwo
 \else \expandafter \@secondoftwo
 \fi
}%
\providecommand \@ifx [1]{%
 \ifx #1\expandafter \@firstoftwo
 \else \expandafter \@secondoftwo
 \fi
}%
\providecommand \natexlab [1]{#1}%
\providecommand \enquote  [1]{``#1''}%
\providecommand \bibnamefont  [1]{#1}%
\providecommand \bibfnamefont [1]{#1}%
\providecommand \citenamefont [1]{#1}%
\providecommand \href@noop [0]{\@secondoftwo}%
\providecommand \href [0]{\begingroup \@sanitize@url \@href}%
\providecommand \@href[1]{\@@startlink{#1}\@@href}%
\providecommand \@@href[1]{\endgroup#1\@@endlink}%
\providecommand \@sanitize@url [0]{\catcode `\\12\catcode `\$12\catcode
  `\&12\catcode `\#12\catcode `\^12\catcode `\_12\catcode `\%12\relax}%
\providecommand \@@startlink[1]{}%
\providecommand \@@endlink[0]{}%
\providecommand \url  [0]{\begingroup\@sanitize@url \@url }%
\providecommand \@url [1]{\endgroup\@href {#1}{\urlprefix }}%
\providecommand \urlprefix  [0]{URL }%
\providecommand \Eprint [0]{\href }%
\providecommand \doibase [0]{https://doi.org/}%
\providecommand \selectlanguage [0]{\@gobble}%
\providecommand \bibinfo  [0]{\@secondoftwo}%
\providecommand \bibfield  [0]{\@secondoftwo}%
\providecommand \translation [1]{[#1]}%
\providecommand \BibitemOpen [0]{}%
\providecommand \bibitemStop [0]{}%
\providecommand \bibitemNoStop [0]{.\EOS\space}%
\providecommand \EOS [0]{\spacefactor3000\relax}%
\providecommand \BibitemShut  [1]{\csname bibitem#1\endcsname}%
\let\auto@bib@innerbib\@empty
%</preamble>
\bibitem [{\citenamefont {K{\"o}hl}(2006)}]{kohl2006}%
  \BibitemOpen
  \bibfield  {author} {\bibinfo {author} {\bibfnamefont {M.}~\bibnamefont
  {K{\"o}hl}},\ }\href {https://doi.org/10.1103/PhysRevA.73.031601} {\bibfield
  {journal} {\bibinfo  {journal} {Physical Review A}\ }\textbf {\bibinfo
  {volume} {73}},\ \bibinfo {pages} {031601} (\bibinfo {year}
  {2006})}\BibitemShut {NoStop}%
\bibitem [{\citenamefont {Zhou}\ and\ \citenamefont {Ho}(2011)}]{zhou2011}%
  \BibitemOpen
  \bibfield  {author} {\bibinfo {author} {\bibfnamefont {Q.}~\bibnamefont
  {Zhou}}\ and\ \bibinfo {author} {\bibfnamefont {T.-L.}\ \bibnamefont {Ho}},\
  }\href {https://doi.org/10.1103/PhysRevLett.106.225301} {\bibfield  {journal}
  {\bibinfo  {journal} {Physical Review Letters}\ }\textbf {\bibinfo {volume}
  {106}},\ \bibinfo {pages} {225301} (\bibinfo {year} {2011})}\BibitemShut
  {NoStop}%
\bibitem [{\citenamefont {Roscilde}(2014)}]{roscilde2014}%
  \BibitemOpen
  \bibfield  {author} {\bibinfo {author} {\bibfnamefont {T.}~\bibnamefont
  {Roscilde}},\ }\href {https://doi.org/10.1103/PhysRevLett.112.110403}
  {\bibfield  {journal} {\bibinfo  {journal} {Physical Review Letters}\
  }\textbf {\bibinfo {volume} {112}},\ \bibinfo {pages} {110403} (\bibinfo
  {year} {2014})}\BibitemShut {NoStop}%
\bibitem [{\citenamefont {McDonald}\ \emph {et~al.}(2015)\citenamefont
  {McDonald}, \citenamefont {McGuyer}, \citenamefont {Iwata},\ and\
  \citenamefont {Zelevinsky}}]{mcdonald2015}%
  \BibitemOpen
  \bibfield  {author} {\bibinfo {author} {\bibfnamefont {M.}~\bibnamefont
  {McDonald}}, \bibinfo {author} {\bibfnamefont {B.~H.}\ \bibnamefont
  {McGuyer}}, \bibinfo {author} {\bibfnamefont {G.~Z.}\ \bibnamefont {Iwata}},\
  and\ \bibinfo {author} {\bibfnamefont {T.}~\bibnamefont {Zelevinsky}},\
  }\href {https://doi.org/10.1103/PhysRevLett.114.023001} {\bibfield  {journal}
  {\bibinfo  {journal} {Physical Review Letters}\ }\textbf {\bibinfo {volume}
  {114}},\ \bibinfo {pages} {023001} (\bibinfo {year} {2015})}\BibitemShut
  {NoStop}%
\bibitem [{\citenamefont {Mitchison}\ \emph {et~al.}(2020)\citenamefont
  {Mitchison}, \citenamefont {Fogarty}, \citenamefont {Guarnieri},
  \citenamefont {Campbell}, \citenamefont {Busch},\ and\ \citenamefont
  {Goold}}]{mitchison2020}%
  \BibitemOpen
  \bibfield  {author} {\bibinfo {author} {\bibfnamefont {M.~T.}\ \bibnamefont
  {Mitchison}}, \bibinfo {author} {\bibfnamefont {T.}~\bibnamefont {Fogarty}},
  \bibinfo {author} {\bibfnamefont {G.}~\bibnamefont {Guarnieri}}, \bibinfo
  {author} {\bibfnamefont {S.}~\bibnamefont {Campbell}}, \bibinfo {author}
  {\bibfnamefont {T.}~\bibnamefont {Busch}},\ and\ \bibinfo {author}
  {\bibfnamefont {J.}~\bibnamefont {Goold}},\ }\href
  {https://doi.org/10.1103/PhysRevLett.125.080402} {\bibfield  {journal}
  {\bibinfo  {journal} {Physical Review Letters}\ }\textbf {\bibinfo {volume}
  {125}},\ \bibinfo {pages} {080402} (\bibinfo {year} {2020})}\BibitemShut
  {NoStop}%
\bibitem [{\citenamefont {Ketterle}\ \emph {et~al.}(1999)\citenamefont
  {Ketterle}, \citenamefont {Durfee},\ and\ \citenamefont
  {{Stamper-Kurn}}}]{ketterle1999}%
  \BibitemOpen
  \bibfield  {author} {\bibinfo {author} {\bibfnamefont {W.}~\bibnamefont
  {Ketterle}}, \bibinfo {author} {\bibfnamefont {D.~S.}\ \bibnamefont
  {Durfee}},\ and\ \bibinfo {author} {\bibfnamefont {D.~M.}\ \bibnamefont
  {{Stamper-Kurn}}},\ }\href {https://doi.org/10.48550/arXiv.cond-mat/9904034}
  {\bibinfo {title} {Making, probing and understanding {{Bose-Einstein}}
  condensates}} (\bibinfo {year} {1999}),\ \Eprint
  {https://arxiv.org/abs/cond-mat/9904034} {arXiv:cond-mat/9904034}
  \BibitemShut {NoStop}%
\bibitem [{\citenamefont {Gemelke}\ \emph {et~al.}(2009)\citenamefont
  {Gemelke}, \citenamefont {Zhang}, \citenamefont {Hung},\ and\ \citenamefont
  {Chin}}]{gemelke2009}%
  \BibitemOpen
  \bibfield  {author} {\bibinfo {author} {\bibfnamefont {N.}~\bibnamefont
  {Gemelke}}, \bibinfo {author} {\bibfnamefont {X.}~\bibnamefont {Zhang}},
  \bibinfo {author} {\bibfnamefont {C.-L.}\ \bibnamefont {Hung}},\ and\
  \bibinfo {author} {\bibfnamefont {C.}~\bibnamefont {Chin}},\ }\href
  {https://doi.org/10.1038/nature08244} {\bibfield  {journal} {\bibinfo
  {journal} {Nature}\ }\textbf {\bibinfo {volume} {460}},\ \bibinfo {pages}
  {995} (\bibinfo {year} {2009})}\BibitemShut {NoStop}%
\bibitem [{\citenamefont {Weld}\ \emph {et~al.}(2009)\citenamefont {Weld},
  \citenamefont {Medley}, \citenamefont {Miyake}, \citenamefont {Hucul},
  \citenamefont {Pritchard},\ and\ \citenamefont {Ketterle}}]{weld2009}%
  \BibitemOpen
  \bibfield  {author} {\bibinfo {author} {\bibfnamefont {D.~M.}\ \bibnamefont
  {Weld}}, \bibinfo {author} {\bibfnamefont {P.}~\bibnamefont {Medley}},
  \bibinfo {author} {\bibfnamefont {H.}~\bibnamefont {Miyake}}, \bibinfo
  {author} {\bibfnamefont {D.}~\bibnamefont {Hucul}}, \bibinfo {author}
  {\bibfnamefont {D.~E.}\ \bibnamefont {Pritchard}},\ and\ \bibinfo {author}
  {\bibfnamefont {W.}~\bibnamefont {Ketterle}},\ }\href
  {https://doi.org/10.1103/PhysRevLett.103.245301} {\bibfield  {journal}
  {\bibinfo  {journal} {Physical Review Letters}\ }\textbf {\bibinfo {volume}
  {103}},\ \bibinfo {pages} {245301} (\bibinfo {year} {2009})}\BibitemShut
  {NoStop}%
\bibitem [{\citenamefont {Sanner}\ \emph {et~al.}(2010)\citenamefont {Sanner},
  \citenamefont {Su}, \citenamefont {Keshet}, \citenamefont {Gommers},
  \citenamefont {Shin}, \citenamefont {Huang},\ and\ \citenamefont
  {Ketterle}}]{sanner2010}%
  \BibitemOpen
  \bibfield  {author} {\bibinfo {author} {\bibfnamefont {C.}~\bibnamefont
  {Sanner}}, \bibinfo {author} {\bibfnamefont {E.~J.}\ \bibnamefont {Su}},
  \bibinfo {author} {\bibfnamefont {A.}~\bibnamefont {Keshet}}, \bibinfo
  {author} {\bibfnamefont {R.}~\bibnamefont {Gommers}}, \bibinfo {author}
  {\bibfnamefont {Y.-i.}\ \bibnamefont {Shin}}, \bibinfo {author}
  {\bibfnamefont {W.}~\bibnamefont {Huang}},\ and\ \bibinfo {author}
  {\bibfnamefont {W.}~\bibnamefont {Ketterle}},\ }\href
  {https://doi.org/10.1103/PhysRevLett.105.040402} {\bibfield  {journal}
  {\bibinfo  {journal} {Physical Review Letters}\ }\textbf {\bibinfo {volume}
  {105}},\ \bibinfo {pages} {040402} (\bibinfo {year} {2010})}\BibitemShut
  {NoStop}%
\bibitem [{\citenamefont {M{\"u}ller}\ \emph {et~al.}(2010)\citenamefont
  {M{\"u}ller}, \citenamefont {Zimmermann}, \citenamefont {Meineke},
  \citenamefont {Brantut}, \citenamefont {Esslinger},\ and\ \citenamefont
  {Moritz}}]{muller2010}%
  \BibitemOpen
  \bibfield  {author} {\bibinfo {author} {\bibfnamefont {T.}~\bibnamefont
  {M{\"u}ller}}, \bibinfo {author} {\bibfnamefont {B.}~\bibnamefont
  {Zimmermann}}, \bibinfo {author} {\bibfnamefont {J.}~\bibnamefont {Meineke}},
  \bibinfo {author} {\bibfnamefont {J.-P.}\ \bibnamefont {Brantut}}, \bibinfo
  {author} {\bibfnamefont {T.}~\bibnamefont {Esslinger}},\ and\ \bibinfo
  {author} {\bibfnamefont {H.}~\bibnamefont {Moritz}},\ }\href
  {https://doi.org/10.1103/PhysRevLett.105.040401} {\bibfield  {journal}
  {\bibinfo  {journal} {Physical Review Letters}\ }\textbf {\bibinfo {volume}
  {105}},\ \bibinfo {pages} {040401} (\bibinfo {year} {2010})}\BibitemShut
  {NoStop}%
\bibitem [{\citenamefont {Ku}\ \emph {et~al.}(2012)\citenamefont {Ku},
  \citenamefont {Sommer}, \citenamefont {Cheuk},\ and\ \citenamefont
  {Zwierlein}}]{ku2012}%
  \BibitemOpen
  \bibfield  {author} {\bibinfo {author} {\bibfnamefont {M.~J.~H.}\
  \bibnamefont {Ku}}, \bibinfo {author} {\bibfnamefont {A.~T.}\ \bibnamefont
  {Sommer}}, \bibinfo {author} {\bibfnamefont {L.~W.}\ \bibnamefont {Cheuk}},\
  and\ \bibinfo {author} {\bibfnamefont {M.~W.}\ \bibnamefont {Zwierlein}},\
  }\href {https://doi.org/10.1126/science.1214987} {\bibfield  {journal}
  {\bibinfo  {journal} {Science}\ }\textbf {\bibinfo {volume} {335}},\ \bibinfo
  {pages} {563} (\bibinfo {year} {2012})}\BibitemShut {NoStop}%
\bibitem [{\citenamefont {Desbuquois}\ \emph {et~al.}(2014)\citenamefont
  {Desbuquois}, \citenamefont {Yefsah}, \citenamefont {Chomaz}, \citenamefont
  {Weitenberg}, \citenamefont {Corman}, \citenamefont {Nascimb{\`e}ne},\ and\
  \citenamefont {Dalibard}}]{desbuquois2014}%
  \BibitemOpen
  \bibfield  {author} {\bibinfo {author} {\bibfnamefont {R.}~\bibnamefont
  {Desbuquois}}, \bibinfo {author} {\bibfnamefont {T.}~\bibnamefont {Yefsah}},
  \bibinfo {author} {\bibfnamefont {L.}~\bibnamefont {Chomaz}}, \bibinfo
  {author} {\bibfnamefont {C.}~\bibnamefont {Weitenberg}}, \bibinfo {author}
  {\bibfnamefont {L.}~\bibnamefont {Corman}}, \bibinfo {author} {\bibfnamefont
  {S.}~\bibnamefont {Nascimb{\`e}ne}},\ and\ \bibinfo {author} {\bibfnamefont
  {J.}~\bibnamefont {Dalibard}},\ }\href
  {https://doi.org/10.1103/PhysRevLett.113.020404} {\bibfield  {journal}
  {\bibinfo  {journal} {Physical Review Letters}\ }\textbf {\bibinfo {volume}
  {113}},\ \bibinfo {pages} {020404} (\bibinfo {year} {2014})}\BibitemShut
  {NoStop}%
\bibitem [{\citenamefont {Tobias}\ \emph {et~al.}(2020)\citenamefont {Tobias},
  \citenamefont {Matsuda}, \citenamefont {Valtolina}, \citenamefont {De~Marco},
  \citenamefont {Li},\ and\ \citenamefont {Ye}}]{tobias2020}%
  \BibitemOpen
  \bibfield  {author} {\bibinfo {author} {\bibfnamefont {W.~G.}\ \bibnamefont
  {Tobias}}, \bibinfo {author} {\bibfnamefont {K.}~\bibnamefont {Matsuda}},
  \bibinfo {author} {\bibfnamefont {G.}~\bibnamefont {Valtolina}}, \bibinfo
  {author} {\bibfnamefont {L.}~\bibnamefont {De~Marco}}, \bibinfo {author}
  {\bibfnamefont {J.-R.}\ \bibnamefont {Li}},\ and\ \bibinfo {author}
  {\bibfnamefont {J.}~\bibnamefont {Ye}},\ }\href
  {https://doi.org/10.1103/PhysRevLett.124.033401} {\bibfield  {journal}
  {\bibinfo  {journal} {Physical Review Letters}\ }\textbf {\bibinfo {volume}
  {124}},\ \bibinfo {pages} {033401} (\bibinfo {year} {2020})}\BibitemShut
  {NoStop}%
\bibitem [{\citenamefont {Yan}\ \emph {et~al.}(2019)\citenamefont {Yan},
  \citenamefont {Patel}, \citenamefont {Mukherjee}, \citenamefont {Fletcher},
  \citenamefont {Struck},\ and\ \citenamefont {Zwierlein}}]{yan2019}%
  \BibitemOpen
  \bibfield  {author} {\bibinfo {author} {\bibfnamefont {Z.}~\bibnamefont
  {Yan}}, \bibinfo {author} {\bibfnamefont {P.~B.}\ \bibnamefont {Patel}},
  \bibinfo {author} {\bibfnamefont {B.}~\bibnamefont {Mukherjee}}, \bibinfo
  {author} {\bibfnamefont {R.~J.}\ \bibnamefont {Fletcher}}, \bibinfo {author}
  {\bibfnamefont {J.}~\bibnamefont {Struck}},\ and\ \bibinfo {author}
  {\bibfnamefont {M.~W.}\ \bibnamefont {Zwierlein}},\ }\href
  {https://doi.org/10.1103/PhysRevLett.122.093401} {\bibfield  {journal}
  {\bibinfo  {journal} {Physical Review Letters}\ }\textbf {\bibinfo {volume}
  {122}},\ \bibinfo {pages} {093401} (\bibinfo {year} {2019})}\BibitemShut
  {NoStop}%
\bibitem [{\citenamefont {Yan}\ \emph {et~al.}(2024)\citenamefont {Yan},
  \citenamefont {Patel}, \citenamefont {Mukherjee}, \citenamefont {Vale},
  \citenamefont {Fletcher},\ and\ \citenamefont {Zwierlein}}]{yan2024}%
  \BibitemOpen
  \bibfield  {author} {\bibinfo {author} {\bibfnamefont {Z.}~\bibnamefont
  {Yan}}, \bibinfo {author} {\bibfnamefont {P.~B.}\ \bibnamefont {Patel}},
  \bibinfo {author} {\bibfnamefont {B.}~\bibnamefont {Mukherjee}}, \bibinfo
  {author} {\bibfnamefont {C.~J.}\ \bibnamefont {Vale}}, \bibinfo {author}
  {\bibfnamefont {R.~J.}\ \bibnamefont {Fletcher}},\ and\ \bibinfo {author}
  {\bibfnamefont {M.~W.}\ \bibnamefont {Zwierlein}},\ }\href
  {https://doi.org/10.1126/science.adg3430} {\bibfield  {journal} {\bibinfo
  {journal} {Science}\ }\textbf {\bibinfo {volume} {383}},\ \bibinfo {pages}
  {629} (\bibinfo {year} {2024})}\BibitemShut {NoStop}%
\bibitem [{\citenamefont {Hung}\ \emph {et~al.}(2011)\citenamefont {Hung},
  \citenamefont {Zhang}, \citenamefont {Gemelke},\ and\ \citenamefont
  {Chin}}]{hung2011}%
  \BibitemOpen
  \bibfield  {author} {\bibinfo {author} {\bibfnamefont {C.-L.}\ \bibnamefont
  {Hung}}, \bibinfo {author} {\bibfnamefont {X.}~\bibnamefont {Zhang}},
  \bibinfo {author} {\bibfnamefont {N.}~\bibnamefont {Gemelke}},\ and\ \bibinfo
  {author} {\bibfnamefont {C.}~\bibnamefont {Chin}},\ }\href
  {https://doi.org/10.1038/nature09722} {\bibfield  {journal} {\bibinfo
  {journal} {Nature}\ }\textbf {\bibinfo {volume} {470}},\ \bibinfo {pages}
  {236} (\bibinfo {year} {2011})}\BibitemShut {NoStop}%
\bibitem [{\citenamefont {Yefsah}\ \emph {et~al.}(2011)\citenamefont {Yefsah},
  \citenamefont {Desbuquois}, \citenamefont {Chomaz}, \citenamefont
  {G{\"u}nter},\ and\ \citenamefont {Dalibard}}]{yefsah2011}%
  \BibitemOpen
  \bibfield  {author} {\bibinfo {author} {\bibfnamefont {T.}~\bibnamefont
  {Yefsah}}, \bibinfo {author} {\bibfnamefont {R.}~\bibnamefont {Desbuquois}},
  \bibinfo {author} {\bibfnamefont {L.}~\bibnamefont {Chomaz}}, \bibinfo
  {author} {\bibfnamefont {K.~J.}\ \bibnamefont {G{\"u}nter}},\ and\ \bibinfo
  {author} {\bibfnamefont {J.}~\bibnamefont {Dalibard}},\ }\href
  {https://doi.org/10.1103/PhysRevLett.107.130401} {\bibfield  {journal}
  {\bibinfo  {journal} {Physical Review Letters}\ }\textbf {\bibinfo {volume}
  {107}},\ \bibinfo {pages} {130401} (\bibinfo {year} {2011})}\BibitemShut
  {NoStop}%
\bibitem [{\citenamefont {Yefsah}(2011)}]{yefsah2011a}%
  \BibitemOpen
  \bibfield  {author} {\bibinfo {author} {\bibfnamefont {T.}~\bibnamefont
  {Yefsah}},\ }\emph {\bibinfo {title} {{Thermodynamique du gaz de Bose {\`a}
  deux dimensions}}},\ \href@noop {} {Ph.D. thesis},\ \bibinfo  {school}
  {Universit{\'e} Pierre et Marie Curie - Paris VI} (\bibinfo {year}
  {2011})\BibitemShut {NoStop}%
\bibitem [{\citenamefont {Drewes}\ \emph {et~al.}(2016)\citenamefont {Drewes},
  \citenamefont {Cocchi}, \citenamefont {Miller}, \citenamefont {Chan},
  \citenamefont {Pertot}, \citenamefont {Brennecke},\ and\ \citenamefont
  {K{\"o}hl}}]{drewes2016}%
  \BibitemOpen
  \bibfield  {author} {\bibinfo {author} {\bibfnamefont {J.~H.}\ \bibnamefont
  {Drewes}}, \bibinfo {author} {\bibfnamefont {E.}~\bibnamefont {Cocchi}},
  \bibinfo {author} {\bibfnamefont {L.~A.}\ \bibnamefont {Miller}}, \bibinfo
  {author} {\bibfnamefont {C.~F.}\ \bibnamefont {Chan}}, \bibinfo {author}
  {\bibfnamefont {D.}~\bibnamefont {Pertot}}, \bibinfo {author} {\bibfnamefont
  {F.}~\bibnamefont {Brennecke}},\ and\ \bibinfo {author} {\bibfnamefont
  {M.}~\bibnamefont {K{\"o}hl}},\ }\href
  {https://doi.org/10.1103/PhysRevLett.117.135301} {\bibfield  {journal}
  {\bibinfo  {journal} {Physical Review Letters}\ }\textbf {\bibinfo {volume}
  {117}},\ \bibinfo {pages} {135301} (\bibinfo {year} {2016})}\BibitemShut
  {NoStop}%
\bibitem [{\citenamefont {Pasqualetti}\ \emph {et~al.}(2024)\citenamefont
  {Pasqualetti}, \citenamefont {Bettermann}, \citenamefont {Darkwah~Oppong},
  \citenamefont {{Ibarra-Garc{\'i}a-Padilla}}, \citenamefont {Dasgupta},
  \citenamefont {Scalettar}, \citenamefont {Hazzard}, \citenamefont {Bloch},\
  and\ \citenamefont {F{\"o}lling}}]{pasqualetti2024}%
  \BibitemOpen
  \bibfield  {author} {\bibinfo {author} {\bibfnamefont {G.}~\bibnamefont
  {Pasqualetti}}, \bibinfo {author} {\bibfnamefont {O.}~\bibnamefont
  {Bettermann}}, \bibinfo {author} {\bibfnamefont {N.}~\bibnamefont
  {Darkwah~Oppong}}, \bibinfo {author} {\bibfnamefont {E.}~\bibnamefont
  {{Ibarra-Garc{\'i}a-Padilla}}}, \bibinfo {author} {\bibfnamefont
  {S.}~\bibnamefont {Dasgupta}}, \bibinfo {author} {\bibfnamefont {R.~T.}\
  \bibnamefont {Scalettar}}, \bibinfo {author} {\bibfnamefont {K.~R.~A.}\
  \bibnamefont {Hazzard}}, \bibinfo {author} {\bibfnamefont {I.}~\bibnamefont
  {Bloch}},\ and\ \bibinfo {author} {\bibfnamefont {S.}~\bibnamefont
  {F{\"o}lling}},\ }\href {https://doi.org/10.1103/PhysRevLett.132.083401}
  {\bibfield  {journal} {\bibinfo  {journal} {Physical Review Letters}\
  }\textbf {\bibinfo {volume} {132}},\ \bibinfo {pages} {083401} (\bibinfo
  {year} {2024})}\BibitemShut {NoStop}%
\bibitem [{\citenamefont {Hartke}\ \emph {et~al.}(2020)\citenamefont {Hartke},
  \citenamefont {Oreg}, \citenamefont {Jia},\ and\ \citenamefont
  {Zwierlein}}]{hartke2020}%
  \BibitemOpen
  \bibfield  {author} {\bibinfo {author} {\bibfnamefont {T.}~\bibnamefont
  {Hartke}}, \bibinfo {author} {\bibfnamefont {B.}~\bibnamefont {Oreg}},
  \bibinfo {author} {\bibfnamefont {N.}~\bibnamefont {Jia}},\ and\ \bibinfo
  {author} {\bibfnamefont {M.}~\bibnamefont {Zwierlein}},\ }\href
  {https://doi.org/10.1103/PhysRevLett.125.113601} {\bibfield  {journal}
  {\bibinfo  {journal} {Physical Review Letters}\ }\textbf {\bibinfo {volume}
  {125}},\ \bibinfo {pages} {113601} (\bibinfo {year} {2020})}\BibitemShut
  {NoStop}%
\bibitem [{\citenamefont {Gring}\ \emph {et~al.}(2012)\citenamefont {Gring},
  \citenamefont {Kuhnert}, \citenamefont {Langen}, \citenamefont {Kitagawa},
  \citenamefont {Rauer}, \citenamefont {Schreitl}, \citenamefont {Mazets},
  \citenamefont {Smith}, \citenamefont {Demler},\ and\ \citenamefont
  {Schmiedmayer}}]{gring2012}%
  \BibitemOpen
  \bibfield  {author} {\bibinfo {author} {\bibfnamefont {M.}~\bibnamefont
  {Gring}}, \bibinfo {author} {\bibfnamefont {M.}~\bibnamefont {Kuhnert}},
  \bibinfo {author} {\bibfnamefont {T.}~\bibnamefont {Langen}}, \bibinfo
  {author} {\bibfnamefont {T.}~\bibnamefont {Kitagawa}}, \bibinfo {author}
  {\bibfnamefont {B.}~\bibnamefont {Rauer}}, \bibinfo {author} {\bibfnamefont
  {M.}~\bibnamefont {Schreitl}}, \bibinfo {author} {\bibfnamefont
  {I.}~\bibnamefont {Mazets}}, \bibinfo {author} {\bibfnamefont {D.~A.}\
  \bibnamefont {Smith}}, \bibinfo {author} {\bibfnamefont {E.}~\bibnamefont
  {Demler}},\ and\ \bibinfo {author} {\bibfnamefont {J.}~\bibnamefont
  {Schmiedmayer}},\ }\href {https://doi.org/10.1126/science.1224953} {\bibfield
   {journal} {\bibinfo  {journal} {Science}\ }\textbf {\bibinfo {volume}
  {337}},\ \bibinfo {pages} {1318} (\bibinfo {year} {2012})}\BibitemShut
  {NoStop}%
\bibitem [{\citenamefont {Trotzky}\ \emph {et~al.}(2012)\citenamefont
  {Trotzky}, \citenamefont {Chen}, \citenamefont {Flesch}, \citenamefont
  {McCulloch}, \citenamefont {Schollw{\"o}ck}, \citenamefont {Eisert},\ and\
  \citenamefont {Bloch}}]{trotzky2012}%
  \BibitemOpen
  \bibfield  {author} {\bibinfo {author} {\bibfnamefont {S.}~\bibnamefont
  {Trotzky}}, \bibinfo {author} {\bibfnamefont {Y.-A.}\ \bibnamefont {Chen}},
  \bibinfo {author} {\bibfnamefont {A.}~\bibnamefont {Flesch}}, \bibinfo
  {author} {\bibfnamefont {I.~P.}\ \bibnamefont {McCulloch}}, \bibinfo {author}
  {\bibfnamefont {U.}~\bibnamefont {Schollw{\"o}ck}}, \bibinfo {author}
  {\bibfnamefont {J.}~\bibnamefont {Eisert}},\ and\ \bibinfo {author}
  {\bibfnamefont {I.}~\bibnamefont {Bloch}},\ }\href
  {https://doi.org/10.1038/nphys2232} {\bibfield  {journal} {\bibinfo
  {journal} {Nature Physics}\ }\textbf {\bibinfo {volume} {8}},\ \bibinfo
  {pages} {325} (\bibinfo {year} {2012})}\BibitemShut {NoStop}%
\bibitem [{\citenamefont {Cheneau}\ \emph {et~al.}(2012)\citenamefont
  {Cheneau}, \citenamefont {Barmettler}, \citenamefont {Poletti}, \citenamefont
  {Endres}, \citenamefont {Schau{\ss}}, \citenamefont {Fukuhara}, \citenamefont
  {Gross}, \citenamefont {Bloch}, \citenamefont {Kollath},\ and\ \citenamefont
  {Kuhr}}]{cheneau2012}%
  \BibitemOpen
  \bibfield  {author} {\bibinfo {author} {\bibfnamefont {M.}~\bibnamefont
  {Cheneau}}, \bibinfo {author} {\bibfnamefont {P.}~\bibnamefont {Barmettler}},
  \bibinfo {author} {\bibfnamefont {D.}~\bibnamefont {Poletti}}, \bibinfo
  {author} {\bibfnamefont {M.}~\bibnamefont {Endres}}, \bibinfo {author}
  {\bibfnamefont {P.}~\bibnamefont {Schau{\ss}}}, \bibinfo {author}
  {\bibfnamefont {T.}~\bibnamefont {Fukuhara}}, \bibinfo {author}
  {\bibfnamefont {C.}~\bibnamefont {Gross}}, \bibinfo {author} {\bibfnamefont
  {I.}~\bibnamefont {Bloch}}, \bibinfo {author} {\bibfnamefont
  {C.}~\bibnamefont {Kollath}},\ and\ \bibinfo {author} {\bibfnamefont
  {S.}~\bibnamefont {Kuhr}},\ }\href {https://doi.org/10.1038/nature10748}
  {\bibfield  {journal} {\bibinfo  {journal} {Nature}\ }\textbf {\bibinfo
  {volume} {481}},\ \bibinfo {pages} {484} (\bibinfo {year}
  {2012})}\BibitemShut {NoStop}%
\bibitem [{\citenamefont {Eigen}\ \emph {et~al.}(2018)\citenamefont {Eigen},
  \citenamefont {Glidden}, \citenamefont {Lopes}, \citenamefont {Cornell},
  \citenamefont {Smith},\ and\ \citenamefont {Hadzibabic}}]{eigen2018}%
  \BibitemOpen
  \bibfield  {author} {\bibinfo {author} {\bibfnamefont {C.}~\bibnamefont
  {Eigen}}, \bibinfo {author} {\bibfnamefont {J.~A.~P.}\ \bibnamefont
  {Glidden}}, \bibinfo {author} {\bibfnamefont {R.}~\bibnamefont {Lopes}},
  \bibinfo {author} {\bibfnamefont {E.~A.}\ \bibnamefont {Cornell}}, \bibinfo
  {author} {\bibfnamefont {R.~P.}\ \bibnamefont {Smith}},\ and\ \bibinfo
  {author} {\bibfnamefont {Z.}~\bibnamefont {Hadzibabic}},\ }\href
  {https://doi.org/10.1038/s41586-018-0674-1} {\bibfield  {journal} {\bibinfo
  {journal} {Nature}\ }\textbf {\bibinfo {volume} {563}},\ \bibinfo {pages}
  {221} (\bibinfo {year} {2018})}\BibitemShut {NoStop}%
\bibitem [{\citenamefont {Kim}\ and\ \citenamefont {Huse}(2012)}]{kim2012}%
  \BibitemOpen
  \bibfield  {author} {\bibinfo {author} {\bibfnamefont {H.}~\bibnamefont
  {Kim}}\ and\ \bibinfo {author} {\bibfnamefont {D.~A.}\ \bibnamefont {Huse}},\
  }\href {https://doi.org/10.1103/PhysRevA.86.053607} {\bibfield  {journal}
  {\bibinfo  {journal} {Physical Review A}\ }\textbf {\bibinfo {volume} {86}},\
  \bibinfo {pages} {053607} (\bibinfo {year} {2012})}\BibitemShut {NoStop}%
\bibitem [{\citenamefont {Samland}\ \emph {et~al.}(2024)\citenamefont
  {Samland}, \citenamefont {Wurz}, \citenamefont {Gall},\ and\ \citenamefont
  {K{\"o}hl}}]{samland2024}%
  \BibitemOpen
  \bibfield  {author} {\bibinfo {author} {\bibfnamefont {J.}~\bibnamefont
  {Samland}}, \bibinfo {author} {\bibfnamefont {N.}~\bibnamefont {Wurz}},
  \bibinfo {author} {\bibfnamefont {M.}~\bibnamefont {Gall}},\ and\ \bibinfo
  {author} {\bibfnamefont {M.}~\bibnamefont {K{\"o}hl}},\ }\href@noop {}
  {\bibinfo {title} {Thermodynamics and density fluctuations in a bilayer
  {{Hubbard}} system of ultracold atoms}} (\bibinfo {year} {2024}),\ \Eprint
  {https://arxiv.org/abs/2407.11863} {arXiv:2407.11863} \BibitemShut {NoStop}%
\bibitem [{\citenamefont {Castin}(2007)}]{castin2007}%
  \BibitemOpen
  \bibfield  {author} {\bibinfo {author} {\bibfnamefont {Y.}~\bibnamefont
  {Castin}},\ }in\ \href {https://doi.org/10.3254/978-1-58603-846-5-289} {\emph
  {\bibinfo {booktitle} {Lecture Notes of the 2006 {{Varenna Enrico Fermi
  School}} on {{Fermi}} Gases}}},\ \bibinfo {editor} {edited by\ \bibinfo
  {editor} {\bibfnamefont {M.}~\bibnamefont {Inguscio}}, \bibinfo {editor}
  {\bibfnamefont {W.}~\bibnamefont {Ketterle}},\ and\ \bibinfo {editor}
  {\bibfnamefont {C.}~\bibnamefont {Salomon}}}\ (\bibinfo  {publisher} {IOS
  Press},\ \bibinfo {year} {2007})\ pp.\ \bibinfo {pages}
  {289--349}\BibitemShut {NoStop}%
\bibitem [{\citenamefont {Verstraten}\ \emph {et~al.}(2024)\citenamefont
  {Verstraten}, \citenamefont {Dai}, \citenamefont {Dixmerias}, \citenamefont
  {Peaudecerf}, \citenamefont {{de Jongh}},\ and\ \citenamefont
  {Yefsah}}]{verstraten2024}%
  \BibitemOpen
  \bibfield  {author} {\bibinfo {author} {\bibfnamefont {J.}~\bibnamefont
  {Verstraten}}, \bibinfo {author} {\bibfnamefont {K.}~\bibnamefont {Dai}},
  \bibinfo {author} {\bibfnamefont {M.}~\bibnamefont {Dixmerias}}, \bibinfo
  {author} {\bibfnamefont {B.}~\bibnamefont {Peaudecerf}}, \bibinfo {author}
  {\bibfnamefont {T.}~\bibnamefont {{de Jongh}}},\ and\ \bibinfo {author}
  {\bibfnamefont {T.}~\bibnamefont {Yefsah}},\ }\href
  {https://doi.org/10.48550/arXiv.2404.05699} {\bibinfo {title} {In-situ
  {{Imaging}} of a {{Single-Atom Wave Packet}} in {{Continuous Space}}}}
  (\bibinfo {year} {2024}),\ \Eprint {https://arxiv.org/abs/2404.05699}
  {arXiv:2404.05699} \BibitemShut {NoStop}%
\bibitem [{\citenamefont {de~Jongh}\ \emph {et~al.}(2024)\citenamefont
  {de~Jongh}, \citenamefont {Verstraten}, \citenamefont {Dixmerias},
  \citenamefont {Daix}, \citenamefont {Peaudecerf},\ and\ \citenamefont
  {Yefsah}}]{jongh2024}%
  \BibitemOpen
  \bibfield  {author} {\bibinfo {author} {\bibfnamefont {T.}~\bibnamefont
  {de~Jongh}}, \bibinfo {author} {\bibfnamefont {J.}~\bibnamefont
  {Verstraten}}, \bibinfo {author} {\bibfnamefont {M.}~\bibnamefont
  {Dixmerias}}, \bibinfo {author} {\bibfnamefont {C.}~\bibnamefont {Daix}},
  \bibinfo {author} {\bibfnamefont {B.}~\bibnamefont {Peaudecerf}},\ and\
  \bibinfo {author} {\bibfnamefont {T.}~\bibnamefont {Yefsah}},\ }\href
  {https://doi.org/10.48550/arXiv.2411.08776} {\bibinfo {title} {Quantum {{Gas
  Microscopy}} of {{Fermions}} in the {{Continuum}}}} (\bibinfo {year}
  {2024}),\ \Eprint {https://arxiv.org/abs/2411.08776} {arXiv:2411.08776}
  \BibitemShut {NoStop}%
\bibitem [{Note1()}]{Note1}%
  \BibitemOpen
  \bibinfo {note} {The symbol $\protect \hat {}$ is omitted for
  readability.}\BibitemShut {Stop}%
\bibitem [{\citenamefont {Astrakharchik}\ \emph {et~al.}(2007)\citenamefont
  {Astrakharchik}, \citenamefont {Combescot},\ and\ \citenamefont
  {Pitaevskii}}]{astrakharchik2007}%
  \BibitemOpen
  \bibfield  {author} {\bibinfo {author} {\bibfnamefont {G.~E.}\ \bibnamefont
  {Astrakharchik}}, \bibinfo {author} {\bibfnamefont {R.}~\bibnamefont
  {Combescot}},\ and\ \bibinfo {author} {\bibfnamefont {L.~P.}\ \bibnamefont
  {Pitaevskii}},\ }\href {https://doi.org/10.1103/PhysRevA.76.063616}
  {\bibfield  {journal} {\bibinfo  {journal} {Physical Review A}\ }\textbf
  {\bibinfo {volume} {76}},\ \bibinfo {pages} {063616} (\bibinfo {year}
  {2007})}\BibitemShut {NoStop}%
\bibitem [{\citenamefont {Klawunn}\ \emph {et~al.}(2011)\citenamefont
  {Klawunn}, \citenamefont {Recati}, \citenamefont {Pitaevskii},\ and\
  \citenamefont {Stringari}}]{klawunn2011}%
  \BibitemOpen
  \bibfield  {author} {\bibinfo {author} {\bibfnamefont {M.}~\bibnamefont
  {Klawunn}}, \bibinfo {author} {\bibfnamefont {A.}~\bibnamefont {Recati}},
  \bibinfo {author} {\bibfnamefont {L.~P.}\ \bibnamefont {Pitaevskii}},\ and\
  \bibinfo {author} {\bibfnamefont {S.}~\bibnamefont {Stringari}},\ }\href
  {https://doi.org/10.1103/PhysRevA.84.033612} {\bibfield  {journal} {\bibinfo
  {journal} {Physical Review A}\ }\textbf {\bibinfo {volume} {84}},\ \bibinfo
  {pages} {033612} (\bibinfo {year} {2011})}\BibitemShut {NoStop}%
\bibitem [{\citenamefont {Omran}\ \emph {et~al.}(2015)\citenamefont {Omran},
  \citenamefont {Boll}, \citenamefont {Hilker}, \citenamefont {Kleinlein},
  \citenamefont {Salomon}, \citenamefont {Bloch},\ and\ \citenamefont
  {Gross}}]{omran2015}%
  \BibitemOpen
  \bibfield  {author} {\bibinfo {author} {\bibfnamefont {A.}~\bibnamefont
  {Omran}}, \bibinfo {author} {\bibfnamefont {M.}~\bibnamefont {Boll}},
  \bibinfo {author} {\bibfnamefont {T.~A.}\ \bibnamefont {Hilker}}, \bibinfo
  {author} {\bibfnamefont {K.}~\bibnamefont {Kleinlein}}, \bibinfo {author}
  {\bibfnamefont {G.}~\bibnamefont {Salomon}}, \bibinfo {author} {\bibfnamefont
  {I.}~\bibnamefont {Bloch}},\ and\ \bibinfo {author} {\bibfnamefont
  {C.}~\bibnamefont {Gross}},\ }\href
  {https://doi.org/10.1103/PhysRevLett.115.263001} {\bibfield  {journal}
  {\bibinfo  {journal} {Physical Review Letters}\ }\textbf {\bibinfo {volume}
  {115}},\ \bibinfo {pages} {263001} (\bibinfo {year} {2015})}\BibitemShut
  {NoStop}%
\bibitem [{Note2()}]{Note2}%
  \BibitemOpen
  \bibinfo {note} {Additional details can be found in the Supplementary
  Materials, which includes Ref. \cite {jin2024}.}\BibitemShut {Stop}%
\bibitem [{\citenamefont {Fr{\'e}rot}\ and\ \citenamefont
  {Roscilde}(2016)}]{frerot2016}%
  \BibitemOpen
  \bibfield  {author} {\bibinfo {author} {\bibfnamefont {I.}~\bibnamefont
  {Fr{\'e}rot}}\ and\ \bibinfo {author} {\bibfnamefont {T.}~\bibnamefont
  {Roscilde}},\ }\href {https://doi.org/10.1103/PhysRevB.94.075121} {\bibfield
  {journal} {\bibinfo  {journal} {Physical Review B}\ }\textbf {\bibinfo
  {volume} {94}},\ \bibinfo {pages} {075121} (\bibinfo {year}
  {2016})}\BibitemShut {NoStop}%
\bibitem [{\citenamefont {Gioev}\ and\ \citenamefont
  {Klich}(2006)}]{gioev2006}%
  \BibitemOpen
  \bibfield  {author} {\bibinfo {author} {\bibfnamefont {D.}~\bibnamefont
  {Gioev}}\ and\ \bibinfo {author} {\bibfnamefont {I.}~\bibnamefont {Klich}},\
  }\href {https://doi.org/10.1103/PhysRevLett.96.100503} {\bibfield  {journal}
  {\bibinfo  {journal} {Physical Review Letters}\ }\textbf {\bibinfo {volume}
  {96}},\ \bibinfo {pages} {100503} (\bibinfo {year} {2006})}\BibitemShut
  {NoStop}%
\bibitem [{\citenamefont {Smith}\ \emph {et~al.}(2021)\citenamefont {Smith},
  \citenamefont {Le~Doussal}, \citenamefont {Majumdar},\ and\ \citenamefont
  {Schehr}}]{smith2021}%
  \BibitemOpen
  \bibfield  {author} {\bibinfo {author} {\bibfnamefont {N.~R.}\ \bibnamefont
  {Smith}}, \bibinfo {author} {\bibfnamefont {P.}~\bibnamefont {Le~Doussal}},
  \bibinfo {author} {\bibfnamefont {S.~N.}\ \bibnamefont {Majumdar}},\ and\
  \bibinfo {author} {\bibfnamefont {G.}~\bibnamefont {Schehr}},\ }\href
  {https://doi.org/10.1103/PhysRevE.103.L030105} {\bibfield  {journal}
  {\bibinfo  {journal} {Physical Review E}\ }\textbf {\bibinfo {volume}
  {103}},\ \bibinfo {pages} {L030105} (\bibinfo {year} {2021})}\BibitemShut
  {NoStop}%
\bibitem [{\citenamefont {Wolf}(2006)}]{wolf2006}%
  \BibitemOpen
  \bibfield  {author} {\bibinfo {author} {\bibfnamefont {M.~M.}\ \bibnamefont
  {Wolf}},\ }\href {https://doi.org/10.1103/PhysRevLett.96.010404} {\bibfield
  {journal} {\bibinfo  {journal} {Physical Review Letters}\ }\textbf {\bibinfo
  {volume} {96}},\ \bibinfo {pages} {010404} (\bibinfo {year}
  {2006})}\BibitemShut {NoStop}%
\bibitem [{\citenamefont {Song}\ \emph {et~al.}(2012)\citenamefont {Song},
  \citenamefont {Rachel}, \citenamefont {Flindt}, \citenamefont {Klich},
  \citenamefont {Laflorencie},\ and\ \citenamefont {Le~Hur}}]{song2012}%
  \BibitemOpen
  \bibfield  {author} {\bibinfo {author} {\bibfnamefont {H.~F.}\ \bibnamefont
  {Song}}, \bibinfo {author} {\bibfnamefont {S.}~\bibnamefont {Rachel}},
  \bibinfo {author} {\bibfnamefont {C.}~\bibnamefont {Flindt}}, \bibinfo
  {author} {\bibfnamefont {I.}~\bibnamefont {Klich}}, \bibinfo {author}
  {\bibfnamefont {N.}~\bibnamefont {Laflorencie}},\ and\ \bibinfo {author}
  {\bibfnamefont {K.}~\bibnamefont {Le~Hur}},\ }\href
  {https://doi.org/10.1103/PhysRevB.85.035409} {\bibfield  {journal} {\bibinfo
  {journal} {Physical Review B}\ }\textbf {\bibinfo {volume} {85}},\ \bibinfo
  {pages} {035409} (\bibinfo {year} {2012})}\BibitemShut {NoStop}%
\bibitem [{\citenamefont {Calabrese}\ \emph {et~al.}(2012)\citenamefont
  {Calabrese}, \citenamefont {Mintchev},\ and\ \citenamefont
  {Vicari}}]{calabrese2012}%
  \BibitemOpen
  \bibfield  {author} {\bibinfo {author} {\bibfnamefont {P.}~\bibnamefont
  {Calabrese}}, \bibinfo {author} {\bibfnamefont {M.}~\bibnamefont
  {Mintchev}},\ and\ \bibinfo {author} {\bibfnamefont {E.}~\bibnamefont
  {Vicari}},\ }\href {https://doi.org/10.1209/0295-5075/98/20003} {\bibfield
  {journal} {\bibinfo  {journal} {EPL (Europhysics Letters)}\ }\textbf
  {\bibinfo {volume} {98}},\ \bibinfo {pages} {20003} (\bibinfo {year}
  {2012})}\BibitemShut {NoStop}%
\bibitem [{\citenamefont {Fr{\'e}rot}\ and\ \citenamefont
  {Roscilde}(2015)}]{frerot2015}%
  \BibitemOpen
  \bibfield  {author} {\bibinfo {author} {\bibfnamefont {I.}~\bibnamefont
  {Fr{\'e}rot}}\ and\ \bibinfo {author} {\bibfnamefont {T.}~\bibnamefont
  {Roscilde}},\ }\href {https://doi.org/10.1103/PhysRevB.92.115129} {\bibfield
  {journal} {\bibinfo  {journal} {Physical Review B}\ }\textbf {\bibinfo
  {volume} {92}},\ \bibinfo {pages} {115129} (\bibinfo {year}
  {2015})}\BibitemShut {NoStop}%
\bibitem [{\citenamefont {Zabrodin}\ and\ \citenamefont
  {Ovchinnikov}(1985)}]{zabrodin1985}%
  \BibitemOpen
  \bibfield  {author} {\bibinfo {author} {\bibfnamefont {A.~V.}\ \bibnamefont
  {Zabrodin}}\ and\ \bibinfo {author} {\bibfnamefont {A.~A.}\ \bibnamefont
  {Ovchinnikov}},\ }\href@noop {} {\  (\bibinfo {year} {1985})}\BibitemShut
  {NoStop}%
\bibitem [{\citenamefont {Berkovich}\ and\ \citenamefont
  {Lowenstein}(1987)}]{berkovich1987}%
  \BibitemOpen
  \bibfield  {author} {\bibinfo {author} {\bibfnamefont {A.}~\bibnamefont
  {Berkovich}}\ and\ \bibinfo {author} {\bibfnamefont {J.~H.}\ \bibnamefont
  {Lowenstein}},\ }\href {https://doi.org/10.1016/0550-3213(87)90329-4}
  {\bibfield  {journal} {\bibinfo  {journal} {Nuclear Physics B}\ }\textbf
  {\bibinfo {volume} {285}},\ \bibinfo {pages} {70} (\bibinfo {year}
  {1987})}\BibitemShut {NoStop}%
\bibitem [{\citenamefont {Leclair}\ \emph {et~al.}(1996)\citenamefont
  {Leclair}, \citenamefont {Lesage}, \citenamefont {Sachdev},\ and\
  \citenamefont {Saleur}}]{leclair1996}%
  \BibitemOpen
  \bibfield  {author} {\bibinfo {author} {\bibfnamefont {A.}~\bibnamefont
  {Leclair}}, \bibinfo {author} {\bibfnamefont {F.}~\bibnamefont {Lesage}},
  \bibinfo {author} {\bibfnamefont {S.}~\bibnamefont {Sachdev}},\ and\ \bibinfo
  {author} {\bibfnamefont {H.}~\bibnamefont {Saleur}},\ }\href
  {https://doi.org/10.1016/S0550-3213(96)00456-7} {\bibfield  {journal}
  {\bibinfo  {journal} {Nuclear Physics B}\ }\textbf {\bibinfo {volume}
  {482}},\ \bibinfo {pages} {579} (\bibinfo {year} {1996})}\BibitemShut
  {NoStop}%
\bibitem [{\citenamefont {LeClair}\ and\ \citenamefont
  {Mussardo}(1999)}]{leclair1999}%
  \BibitemOpen
  \bibfield  {author} {\bibinfo {author} {\bibfnamefont {A.}~\bibnamefont
  {LeClair}}\ and\ \bibinfo {author} {\bibfnamefont {G.}~\bibnamefont
  {Mussardo}},\ }\href {https://doi.org/10.1016/S0550-3213(99)00280-1}
  {\bibfield  {journal} {\bibinfo  {journal} {Nuclear Physics B}\ }\textbf
  {\bibinfo {volume} {552}},\ \bibinfo {pages} {624} (\bibinfo {year}
  {1999})}\BibitemShut {NoStop}%
\bibitem [{\citenamefont {Kheruntsyan}\ \emph {et~al.}(2003)\citenamefont
  {Kheruntsyan}, \citenamefont {Gangardt}, \citenamefont {Drummond},\ and\
  \citenamefont {Shlyapnikov}}]{kheruntsyan2003}%
  \BibitemOpen
  \bibfield  {author} {\bibinfo {author} {\bibfnamefont {K.~V.}\ \bibnamefont
  {Kheruntsyan}}, \bibinfo {author} {\bibfnamefont {D.~M.}\ \bibnamefont
  {Gangardt}}, \bibinfo {author} {\bibfnamefont {P.~D.}\ \bibnamefont
  {Drummond}},\ and\ \bibinfo {author} {\bibfnamefont {G.~V.}\ \bibnamefont
  {Shlyapnikov}},\ }\href {https://doi.org/10.1103/PhysRevLett.91.040403}
  {\bibfield  {journal} {\bibinfo  {journal} {Physical Review Letters}\
  }\textbf {\bibinfo {volume} {91}},\ \bibinfo {pages} {040403} (\bibinfo
  {year} {2003})}\BibitemShut {NoStop}%
\bibitem [{\citenamefont {Caux}\ and\ \citenamefont
  {Calabrese}(2006)}]{caux2006}%
  \BibitemOpen
  \bibfield  {author} {\bibinfo {author} {\bibfnamefont {J.-S.}\ \bibnamefont
  {Caux}}\ and\ \bibinfo {author} {\bibfnamefont {P.}~\bibnamefont
  {Calabrese}},\ }\href {https://doi.org/10.1103/PhysRevA.74.031605} {\bibfield
   {journal} {\bibinfo  {journal} {Physical Review A}\ }\textbf {\bibinfo
  {volume} {74}},\ \bibinfo {pages} {031605} (\bibinfo {year}
  {2006})}\BibitemShut {NoStop}%
\bibitem [{\citenamefont {Dzyaloshinski{\u I}}\ and\ \citenamefont
  {Larkin}(1996)}]{dzyaloshinskii1996}%
  \BibitemOpen
  \bibfield  {author} {\bibinfo {author} {\bibfnamefont {I.~E.}\ \bibnamefont
  {Dzyaloshinski{\u I}}}\ and\ \bibinfo {author} {\bibfnamefont {A.~I.}\
  \bibnamefont {Larkin}},\ }\bibinfo {title} {Correlation functions for a
  one-dimensional {{Fermi}} system with long-range interaction ({{Tomonaga}}
  model)},\ in\ \href {https://doi.org/10.1142/9789814317344_0014} {\emph
  {\bibinfo {booktitle} {World {{Scientific Series}} in 20th {{Century
  Physics}}}}},\ Vol.~\bibinfo {volume} {11}\ (\bibinfo  {publisher} {WORLD
  SCIENTIFIC},\ \bibinfo {year} {1996})\ pp.\ \bibinfo {pages}
  {95--101}\BibitemShut {NoStop}%
\bibitem [{\citenamefont {Ceperley}(1995)}]{ceperley1995}%
  \BibitemOpen
  \bibfield  {author} {\bibinfo {author} {\bibfnamefont {D.~M.}\ \bibnamefont
  {Ceperley}},\ }\href {https://doi.org/10.1103/RevModPhys.67.279} {\bibfield
  {journal} {\bibinfo  {journal} {Reviews of Modern Physics}\ }\textbf
  {\bibinfo {volume} {67}},\ \bibinfo {pages} {279} (\bibinfo {year}
  {1995})}\BibitemShut {NoStop}%
\bibitem [{\citenamefont {Pilati}\ \emph {et~al.}(2005)\citenamefont {Pilati},
  \citenamefont {Boronat}, \citenamefont {Casulleras},\ and\ \citenamefont
  {Giorgini}}]{pilati2005}%
  \BibitemOpen
  \bibfield  {author} {\bibinfo {author} {\bibfnamefont {S.}~\bibnamefont
  {Pilati}}, \bibinfo {author} {\bibfnamefont {J.}~\bibnamefont {Boronat}},
  \bibinfo {author} {\bibfnamefont {J.}~\bibnamefont {Casulleras}},\ and\
  \bibinfo {author} {\bibfnamefont {S.}~\bibnamefont {Giorgini}},\ }\href
  {https://doi.org/10.1103/PhysRevA.71.023605} {\bibfield  {journal} {\bibinfo
  {journal} {Physical Review A}\ }\textbf {\bibinfo {volume} {71}},\ \bibinfo
  {pages} {023605} (\bibinfo {year} {2005})}\BibitemShut {NoStop}%
\bibitem [{\citenamefont {Holzmann}\ and\ \citenamefont
  {Castin}(1999)}]{holzmann1999}%
  \BibitemOpen
  \bibfield  {author} {\bibinfo {author} {\bibfnamefont {M.}~\bibnamefont
  {Holzmann}}\ and\ \bibinfo {author} {\bibfnamefont {Y.}~\bibnamefont
  {Castin}},\ }\href {https://doi.org/10.1007/s100530050586} {\bibfield
  {journal} {\bibinfo  {journal} {The European Physical Journal D - Atomic,
  Molecular, Optical and Plasma Physics}\ }\textbf {\bibinfo {volume} {7}},\
  \bibinfo {pages} {425} (\bibinfo {year} {1999})}\BibitemShut {NoStop}%
\bibitem [{\citenamefont {{Capogrosso-Sansone}}\ \emph
  {et~al.}(2008)\citenamefont {{Capogrosso-Sansone}}, \citenamefont
  {S{\"o}yler}, \citenamefont {Prokof'ev},\ and\ \citenamefont
  {Svistunov}}]{capogrosso-sansone2008}%
  \BibitemOpen
  \bibfield  {author} {\bibinfo {author} {\bibfnamefont {B.}~\bibnamefont
  {{Capogrosso-Sansone}}}, \bibinfo {author} {\bibfnamefont {{\c S}.~G.}\
  \bibnamefont {S{\"o}yler}}, \bibinfo {author} {\bibfnamefont
  {N.}~\bibnamefont {Prokof'ev}},\ and\ \bibinfo {author} {\bibfnamefont
  {B.}~\bibnamefont {Svistunov}},\ }\href
  {https://doi.org/10.1103/PhysRevA.77.015602} {\bibfield  {journal} {\bibinfo
  {journal} {Physical Review A}\ }\textbf {\bibinfo {volume} {77}},\ \bibinfo
  {pages} {015602} (\bibinfo {year} {2008})}\BibitemShut {NoStop}%
\bibitem [{\citenamefont {Caleffi}\ \emph {et~al.}(2020)\citenamefont
  {Caleffi}, \citenamefont {Capone}, \citenamefont {Menotti}, \citenamefont
  {Carusotto},\ and\ \citenamefont {Recati}}]{caleffi2020}%
  \BibitemOpen
  \bibfield  {author} {\bibinfo {author} {\bibfnamefont {F.}~\bibnamefont
  {Caleffi}}, \bibinfo {author} {\bibfnamefont {M.}~\bibnamefont {Capone}},
  \bibinfo {author} {\bibfnamefont {C.}~\bibnamefont {Menotti}}, \bibinfo
  {author} {\bibfnamefont {I.}~\bibnamefont {Carusotto}},\ and\ \bibinfo
  {author} {\bibfnamefont {A.}~\bibnamefont {Recati}},\ }\href
  {https://doi.org/10.1103/PhysRevResearch.2.033276} {\bibfield  {journal}
  {\bibinfo  {journal} {Physical Review Research}\ }\textbf {\bibinfo {volume}
  {2}},\ \bibinfo {pages} {033276} (\bibinfo {year} {2020})}\BibitemShut
  {NoStop}%
\bibitem [{\citenamefont {Batrouni}\ and\ \citenamefont
  {Scalettar}(2000)}]{batrouni2000}%
  \BibitemOpen
  \bibfield  {author} {\bibinfo {author} {\bibfnamefont {G.~G.}\ \bibnamefont
  {Batrouni}}\ and\ \bibinfo {author} {\bibfnamefont {R.~T.}\ \bibnamefont
  {Scalettar}},\ }\href {https://doi.org/10.1103/PhysRevLett.84.1599}
  {\bibfield  {journal} {\bibinfo  {journal} {Physical Review Letters}\
  }\textbf {\bibinfo {volume} {84}},\ \bibinfo {pages} {1599} (\bibinfo {year}
  {2000})}\BibitemShut {NoStop}%
\bibitem [{\citenamefont {{Capogrosso-Sansone}}\ \emph
  {et~al.}(2010)\citenamefont {{Capogrosso-Sansone}}, \citenamefont {Trefzger},
  \citenamefont {Lewenstein}, \citenamefont {Zoller},\ and\ \citenamefont
  {Pupillo}}]{capogrosso-sansone2010}%
  \BibitemOpen
  \bibfield  {author} {\bibinfo {author} {\bibfnamefont {B.}~\bibnamefont
  {{Capogrosso-Sansone}}}, \bibinfo {author} {\bibfnamefont {C.}~\bibnamefont
  {Trefzger}}, \bibinfo {author} {\bibfnamefont {M.}~\bibnamefont
  {Lewenstein}}, \bibinfo {author} {\bibfnamefont {P.}~\bibnamefont {Zoller}},\
  and\ \bibinfo {author} {\bibfnamefont {G.}~\bibnamefont {Pupillo}},\ }\href
  {https://doi.org/10.1103/PhysRevLett.104.125301} {\bibfield  {journal}
  {\bibinfo  {journal} {Physical Review Letters}\ }\textbf {\bibinfo {volume}
  {104}},\ \bibinfo {pages} {125301} (\bibinfo {year} {2010})}\BibitemShut
  {NoStop}%
\bibitem [{\citenamefont {Yamamoto}\ \emph {et~al.}(2012)\citenamefont
  {Yamamoto}, \citenamefont {Masaki},\ and\ \citenamefont
  {Danshita}}]{yamamoto2012}%
  \BibitemOpen
  \bibfield  {author} {\bibinfo {author} {\bibfnamefont {D.}~\bibnamefont
  {Yamamoto}}, \bibinfo {author} {\bibfnamefont {A.}~\bibnamefont {Masaki}},\
  and\ \bibinfo {author} {\bibfnamefont {I.}~\bibnamefont {Danshita}},\ }\href
  {https://doi.org/10.1103/PhysRevB.86.054516} {\bibfield  {journal} {\bibinfo
  {journal} {Physical Review B}\ }\textbf {\bibinfo {volume} {86}},\ \bibinfo
  {pages} {054516} (\bibinfo {year} {2012})}\BibitemShut {NoStop}%
\bibitem [{\citenamefont {Zhang}(1999)}]{zhang1999}%
  \BibitemOpen
  \bibfield  {author} {\bibinfo {author} {\bibfnamefont {S.}~\bibnamefont
  {Zhang}},\ }\href {https://doi.org/10.1103/PhysRevLett.83.2777} {\bibfield
  {journal} {\bibinfo  {journal} {Physical Review Letters}\ }\textbf {\bibinfo
  {volume} {83}},\ \bibinfo {pages} {2777} (\bibinfo {year}
  {1999})}\BibitemShut {NoStop}%
\bibitem [{\citenamefont {Varney}\ \emph {et~al.}(2009)\citenamefont {Varney},
  \citenamefont {Lee}, \citenamefont {Bai}, \citenamefont {Chiesa},
  \citenamefont {Jarrell},\ and\ \citenamefont {Scalettar}}]{varney2009}%
  \BibitemOpen
  \bibfield  {author} {\bibinfo {author} {\bibfnamefont {C.~N.}\ \bibnamefont
  {Varney}}, \bibinfo {author} {\bibfnamefont {C.-R.}\ \bibnamefont {Lee}},
  \bibinfo {author} {\bibfnamefont {Z.~J.}\ \bibnamefont {Bai}}, \bibinfo
  {author} {\bibfnamefont {S.}~\bibnamefont {Chiesa}}, \bibinfo {author}
  {\bibfnamefont {M.}~\bibnamefont {Jarrell}},\ and\ \bibinfo {author}
  {\bibfnamefont {R.~T.}\ \bibnamefont {Scalettar}},\ }\href
  {https://doi.org/10.1103/PhysRevB.80.075116} {\bibfield  {journal} {\bibinfo
  {journal} {Physical Review B}\ }\textbf {\bibinfo {volume} {80}},\ \bibinfo
  {pages} {075116} (\bibinfo {year} {2009})}\BibitemShut {NoStop}%
\bibitem [{\citenamefont {Paiva}\ \emph {et~al.}(2010)\citenamefont {Paiva},
  \citenamefont {Scalettar}, \citenamefont {Randeria},\ and\ \citenamefont
  {Trivedi}}]{paiva2010}%
  \BibitemOpen
  \bibfield  {author} {\bibinfo {author} {\bibfnamefont {T.}~\bibnamefont
  {Paiva}}, \bibinfo {author} {\bibfnamefont {R.}~\bibnamefont {Scalettar}},
  \bibinfo {author} {\bibfnamefont {M.}~\bibnamefont {Randeria}},\ and\
  \bibinfo {author} {\bibfnamefont {N.}~\bibnamefont {Trivedi}},\ }\href
  {https://doi.org/10.1103/PhysRevLett.104.066406} {\bibfield  {journal}
  {\bibinfo  {journal} {Physical Review Letters}\ }\textbf {\bibinfo {volume}
  {104}},\ \bibinfo {pages} {066406} (\bibinfo {year} {2010})}\BibitemShut
  {NoStop}%
\bibitem [{\citenamefont {Cheuk}\ \emph {et~al.}(2016)\citenamefont {Cheuk},
  \citenamefont {Nichols}, \citenamefont {Lawrence}, \citenamefont {Okan},
  \citenamefont {Zhang}, \citenamefont {Khatami}, \citenamefont {Trivedi},
  \citenamefont {Paiva}, \citenamefont {Rigol},\ and\ \citenamefont
  {Zwierlein}}]{cheuk2016}%
  \BibitemOpen
  \bibfield  {author} {\bibinfo {author} {\bibfnamefont {L.~W.}\ \bibnamefont
  {Cheuk}}, \bibinfo {author} {\bibfnamefont {M.~A.}\ \bibnamefont {Nichols}},
  \bibinfo {author} {\bibfnamefont {K.~R.}\ \bibnamefont {Lawrence}}, \bibinfo
  {author} {\bibfnamefont {M.}~\bibnamefont {Okan}}, \bibinfo {author}
  {\bibfnamefont {H.}~\bibnamefont {Zhang}}, \bibinfo {author} {\bibfnamefont
  {E.}~\bibnamefont {Khatami}}, \bibinfo {author} {\bibfnamefont
  {N.}~\bibnamefont {Trivedi}}, \bibinfo {author} {\bibfnamefont
  {T.}~\bibnamefont {Paiva}}, \bibinfo {author} {\bibfnamefont
  {M.}~\bibnamefont {Rigol}},\ and\ \bibinfo {author} {\bibfnamefont {M.~W.}\
  \bibnamefont {Zwierlein}},\ }\href {https://doi.org/10.1126/science.aag3349}
  {\bibfield  {journal} {\bibinfo  {journal} {Science}\ }\textbf {\bibinfo
  {volume} {353}},\ \bibinfo {pages} {1260} (\bibinfo {year}
  {2016})}\BibitemShut {NoStop}%
\bibitem [{\citenamefont {He}\ \emph {et~al.}(2019)\citenamefont {He},
  \citenamefont {Qin}, \citenamefont {Shi}, \citenamefont {Lu},\ and\
  \citenamefont {Zhang}}]{he2019}%
  \BibitemOpen
  \bibfield  {author} {\bibinfo {author} {\bibfnamefont {Y.-Y.}\ \bibnamefont
  {He}}, \bibinfo {author} {\bibfnamefont {M.}~\bibnamefont {Qin}}, \bibinfo
  {author} {\bibfnamefont {H.}~\bibnamefont {Shi}}, \bibinfo {author}
  {\bibfnamefont {Z.-Y.}\ \bibnamefont {Lu}},\ and\ \bibinfo {author}
  {\bibfnamefont {S.}~\bibnamefont {Zhang}},\ }\href
  {https://doi.org/10.1103/PhysRevB.99.045108} {\bibfield  {journal} {\bibinfo
  {journal} {Physical Review B}\ }\textbf {\bibinfo {volume} {99}},\ \bibinfo
  {pages} {045108} (\bibinfo {year} {2019})}\BibitemShut {NoStop}%
\bibitem [{\citenamefont {Chan}\ \emph {et~al.}(2020)\citenamefont {Chan},
  \citenamefont {Gall}, \citenamefont {Wurz},\ and\ \citenamefont
  {K{\"o}hl}}]{chan2020}%
  \BibitemOpen
  \bibfield  {author} {\bibinfo {author} {\bibfnamefont {C.~F.}\ \bibnamefont
  {Chan}}, \bibinfo {author} {\bibfnamefont {M.}~\bibnamefont {Gall}}, \bibinfo
  {author} {\bibfnamefont {N.}~\bibnamefont {Wurz}},\ and\ \bibinfo {author}
  {\bibfnamefont {M.}~\bibnamefont {K{\"o}hl}},\ }\href
  {https://doi.org/10.1103/PhysRevResearch.2.023210} {\bibfield  {journal}
  {\bibinfo  {journal} {Physical Review Research}\ }\textbf {\bibinfo {volume}
  {2}},\ \bibinfo {pages} {023210} (\bibinfo {year} {2020})}\BibitemShut
  {NoStop}%
\bibitem [{\citenamefont {Pollet}(2012)}]{pollet2012}%
  \BibitemOpen
  \bibfield  {author} {\bibinfo {author} {\bibfnamefont {L.}~\bibnamefont
  {Pollet}},\ }\href {https://doi.org/10.1088/0034-4885/75/9/094501} {\bibfield
   {journal} {\bibinfo  {journal} {Reports on Progress in Physics}\ }\textbf
  {\bibinfo {volume} {75}},\ \bibinfo {pages} {094501} (\bibinfo {year}
  {2012})}\BibitemShut {NoStop}%
\bibitem [{\citenamefont {Chang}\ \emph {et~al.}(2015)\citenamefont {Chang},
  \citenamefont {Gogolenko}, \citenamefont {Perez}, \citenamefont {Bai},\ and\
  \citenamefont {Scalettar}}]{chang2015}%
  \BibitemOpen
  \bibfield  {author} {\bibinfo {author} {\bibfnamefont {C.-C.}\ \bibnamefont
  {Chang}}, \bibinfo {author} {\bibfnamefont {S.}~\bibnamefont {Gogolenko}},
  \bibinfo {author} {\bibfnamefont {J.}~\bibnamefont {Perez}}, \bibinfo
  {author} {\bibfnamefont {Z.}~\bibnamefont {Bai}},\ and\ \bibinfo {author}
  {\bibfnamefont {R.~T.}\ \bibnamefont {Scalettar}},\ }\href
  {https://doi.org/10.1080/14786435.2013.845314} {\bibfield  {journal}
  {\bibinfo  {journal} {Philosophical Magazine}\ }\textbf {\bibinfo {volume}
  {95}},\ \bibinfo {pages} {1260} (\bibinfo {year} {2015})}\BibitemShut
  {NoStop}%
\bibitem [{\citenamefont {Alexandru}\ \emph {et~al.}(2022)\citenamefont
  {Alexandru}, \citenamefont {Ba{\c s}ar}, \citenamefont {Bedaque},\ and\
  \citenamefont {Warrington}}]{alexandru2022}%
  \BibitemOpen
  \bibfield  {author} {\bibinfo {author} {\bibfnamefont {A.}~\bibnamefont
  {Alexandru}}, \bibinfo {author} {\bibfnamefont {G.}~\bibnamefont {Ba{\c
  s}ar}}, \bibinfo {author} {\bibfnamefont {P.~F.}\ \bibnamefont {Bedaque}},\
  and\ \bibinfo {author} {\bibfnamefont {N.~C.}\ \bibnamefont {Warrington}},\
  }\href {https://doi.org/10.1103/RevModPhys.94.015006} {\bibfield  {journal}
  {\bibinfo  {journal} {Reviews of Modern Physics}\ }\textbf {\bibinfo {volume}
  {94}},\ \bibinfo {pages} {015006} (\bibinfo {year} {2022})}\BibitemShut
  {NoStop}%
\bibitem [{\citenamefont {Carlson}\ \emph {et~al.}(2011)\citenamefont
  {Carlson}, \citenamefont {Gandolfi}, \citenamefont {Schmidt},\ and\
  \citenamefont {Zhang}}]{carlson2011}%
  \BibitemOpen
  \bibfield  {author} {\bibinfo {author} {\bibfnamefont {J.}~\bibnamefont
  {Carlson}}, \bibinfo {author} {\bibfnamefont {S.}~\bibnamefont {Gandolfi}},
  \bibinfo {author} {\bibfnamefont {K.~E.}\ \bibnamefont {Schmidt}},\ and\
  \bibinfo {author} {\bibfnamefont {S.}~\bibnamefont {Zhang}},\ }\href
  {https://doi.org/10.1103/PhysRevA.84.061602} {\bibfield  {journal} {\bibinfo
  {journal} {Physical Review A}\ }\textbf {\bibinfo {volume} {84}},\ \bibinfo
  {pages} {061602} (\bibinfo {year} {2011})}\BibitemShut {NoStop}%
\bibitem [{\citenamefont {Van~Houcke}\ \emph {et~al.}(2012)\citenamefont
  {Van~Houcke}, \citenamefont {Werner}, \citenamefont {Kozik}, \citenamefont
  {Prokof'ev}, \citenamefont {Svistunov}, \citenamefont {Ku}, \citenamefont
  {Sommer}, \citenamefont {Cheuk}, \citenamefont {Schirotzek},\ and\
  \citenamefont {Zwierlein}}]{vanhoucke2012}%
  \BibitemOpen
  \bibfield  {author} {\bibinfo {author} {\bibfnamefont {K.}~\bibnamefont
  {Van~Houcke}}, \bibinfo {author} {\bibfnamefont {F.}~\bibnamefont {Werner}},
  \bibinfo {author} {\bibfnamefont {E.}~\bibnamefont {Kozik}}, \bibinfo
  {author} {\bibfnamefont {N.}~\bibnamefont {Prokof'ev}}, \bibinfo {author}
  {\bibfnamefont {B.}~\bibnamefont {Svistunov}}, \bibinfo {author}
  {\bibfnamefont {M.~J.~H.}\ \bibnamefont {Ku}}, \bibinfo {author}
  {\bibfnamefont {A.~T.}\ \bibnamefont {Sommer}}, \bibinfo {author}
  {\bibfnamefont {L.~W.}\ \bibnamefont {Cheuk}}, \bibinfo {author}
  {\bibfnamefont {A.}~\bibnamefont {Schirotzek}},\ and\ \bibinfo {author}
  {\bibfnamefont {M.~W.}\ \bibnamefont {Zwierlein}},\ }\href
  {https://doi.org/10.1038/nphys2273} {\bibfield  {journal} {\bibinfo
  {journal} {Nature Physics}\ }\textbf {\bibinfo {volume} {8}},\ \bibinfo
  {pages} {366} (\bibinfo {year} {2012})}\BibitemShut {NoStop}%
\bibitem [{\citenamefont {Rossi}\ \emph {et~al.}(2018)\citenamefont {Rossi},
  \citenamefont {Ohgoe}, \citenamefont {Van~Houcke},\ and\ \citenamefont
  {Werner}}]{rossi2018}%
  \BibitemOpen
  \bibfield  {author} {\bibinfo {author} {\bibfnamefont {R.}~\bibnamefont
  {Rossi}}, \bibinfo {author} {\bibfnamefont {T.}~\bibnamefont {Ohgoe}},
  \bibinfo {author} {\bibfnamefont {K.}~\bibnamefont {Van~Houcke}},\ and\
  \bibinfo {author} {\bibfnamefont {F.}~\bibnamefont {Werner}},\ }\href
  {https://doi.org/10.1103/PhysRevLett.121.130405} {\bibfield  {journal}
  {\bibinfo  {journal} {Physical Review Letters}\ }\textbf {\bibinfo {volume}
  {121}},\ \bibinfo {pages} {130405} (\bibinfo {year} {2018})}\BibitemShut
  {NoStop}%
\bibitem [{\citenamefont {Shin}\ \emph {et~al.}(2006)\citenamefont {Shin},
  \citenamefont {Zwierlein}, \citenamefont {Schunck}, \citenamefont
  {Schirotzek},\ and\ \citenamefont {Ketterle}}]{Shin2006}%
  \BibitemOpen
  \bibfield  {author} {\bibinfo {author} {\bibfnamefont {Y.}~\bibnamefont
  {Shin}}, \bibinfo {author} {\bibfnamefont {M.~W.}\ \bibnamefont {Zwierlein}},
  \bibinfo {author} {\bibfnamefont {C.~H.}\ \bibnamefont {Schunck}}, \bibinfo
  {author} {\bibfnamefont {A.}~\bibnamefont {Schirotzek}},\ and\ \bibinfo
  {author} {\bibfnamefont {W.}~\bibnamefont {Ketterle}},\ }\href
  {https://doi.org/10.1103/PhysRevLett.97.030401} {\bibfield  {journal}
  {\bibinfo  {journal} {Physical Review Letters}\ }\textbf {\bibinfo {volume}
  {97}},\ \bibinfo {pages} {030401} (\bibinfo {year} {2006})}\BibitemShut
  {NoStop}%
\bibitem [{\citenamefont {Zwierlein}\ \emph {et~al.}(2006)\citenamefont
  {Zwierlein}, \citenamefont {Schirotzek}, \citenamefont {Schunck},\ and\
  \citenamefont {Ketterle}}]{Zwierlein2006}%
  \BibitemOpen
  \bibfield  {author} {\bibinfo {author} {\bibfnamefont {M.~W.}\ \bibnamefont
  {Zwierlein}}, \bibinfo {author} {\bibfnamefont {A.}~\bibnamefont
  {Schirotzek}}, \bibinfo {author} {\bibfnamefont {C.~H.}\ \bibnamefont
  {Schunck}},\ and\ \bibinfo {author} {\bibfnamefont {W.}~\bibnamefont
  {Ketterle}},\ }\href {https://doi.org/10.1126/science.1122318} {\bibfield
  {journal} {\bibinfo  {journal} {Science}\ }\textbf {\bibinfo {volume}
  {311}},\ \bibinfo {pages} {492} (\bibinfo {year} {2006})}\BibitemShut
  {NoStop}%
\bibitem [{\citenamefont {Shin}\ \emph {et~al.}(2008)\citenamefont {Shin},
  \citenamefont {Schunck}, \citenamefont {Schirotzek},\ and\ \citenamefont
  {Ketterle}}]{Shin2008}%
  \BibitemOpen
  \bibfield  {author} {\bibinfo {author} {\bibfnamefont {Y.-i.}\ \bibnamefont
  {Shin}}, \bibinfo {author} {\bibfnamefont {C.~H.}\ \bibnamefont {Schunck}},
  \bibinfo {author} {\bibfnamefont {A.}~\bibnamefont {Schirotzek}},\ and\
  \bibinfo {author} {\bibfnamefont {W.}~\bibnamefont {Ketterle}},\ }\href
  {https://doi.org/10.1038/nature06473} {\bibfield  {journal} {\bibinfo
  {journal} {Nature}\ }\textbf {\bibinfo {volume} {451}},\ \bibinfo {pages}
  {689} (\bibinfo {year} {2008})}\BibitemShut {NoStop}%
\bibitem [{\citenamefont {Mitra}\ \emph {et~al.}(2016)\citenamefont {Mitra},
  \citenamefont {Brown}, \citenamefont {Schau{\ss}}, \citenamefont {Kondov},\
  and\ \citenamefont {Bakr}}]{Mitra2016}%
  \BibitemOpen
  \bibfield  {author} {\bibinfo {author} {\bibfnamefont {D.}~\bibnamefont
  {Mitra}}, \bibinfo {author} {\bibfnamefont {P.~T.}\ \bibnamefont {Brown}},
  \bibinfo {author} {\bibfnamefont {P.}~\bibnamefont {Schau{\ss}}}, \bibinfo
  {author} {\bibfnamefont {S.~S.}\ \bibnamefont {Kondov}},\ and\ \bibinfo
  {author} {\bibfnamefont {W.~S.}\ \bibnamefont {Bakr}},\ }\href
  {https://doi.org/10.1103/PhysRevLett.117.093601} {\bibfield  {journal}
  {\bibinfo  {journal} {Physical Review Letters}\ }\textbf {\bibinfo {volume}
  {117}},\ \bibinfo {pages} {093601} (\bibinfo {year} {2016})}\BibitemShut
  {NoStop}%
\bibitem [{\citenamefont {Mazurenko}\ \emph {et~al.}(2017)\citenamefont
  {Mazurenko}, \citenamefont {Chiu}, \citenamefont {Ji}, \citenamefont
  {Parsons}, \citenamefont {{Kan{\'a}sz-Nagy}}, \citenamefont {Schmidt},
  \citenamefont {Grusdt}, \citenamefont {Demler}, \citenamefont {Greif},\ and\
  \citenamefont {Greiner}}]{mazurenko2017}%
  \BibitemOpen
  \bibfield  {author} {\bibinfo {author} {\bibfnamefont {A.}~\bibnamefont
  {Mazurenko}}, \bibinfo {author} {\bibfnamefont {C.~S.}\ \bibnamefont {Chiu}},
  \bibinfo {author} {\bibfnamefont {G.}~\bibnamefont {Ji}}, \bibinfo {author}
  {\bibfnamefont {M.~F.}\ \bibnamefont {Parsons}}, \bibinfo {author}
  {\bibfnamefont {M.}~\bibnamefont {{Kan{\'a}sz-Nagy}}}, \bibinfo {author}
  {\bibfnamefont {R.}~\bibnamefont {Schmidt}}, \bibinfo {author} {\bibfnamefont
  {F.}~\bibnamefont {Grusdt}}, \bibinfo {author} {\bibfnamefont
  {E.}~\bibnamefont {Demler}}, \bibinfo {author} {\bibfnamefont
  {D.}~\bibnamefont {Greif}},\ and\ \bibinfo {author} {\bibfnamefont
  {M.}~\bibnamefont {Greiner}},\ }\href {https://doi.org/10.1038/nature22362}
  {\bibfield  {journal} {\bibinfo  {journal} {Nature}\ }\textbf {\bibinfo
  {volume} {545}},\ \bibinfo {pages} {462} (\bibinfo {year}
  {2017})}\BibitemShut {NoStop}%
\bibitem [{\citenamefont {Chiu}\ \emph {et~al.}(2018)\citenamefont {Chiu},
  \citenamefont {Ji}, \citenamefont {Mazurenko}, \citenamefont {Greif},\ and\
  \citenamefont {Greiner}}]{chiu2018}%
  \BibitemOpen
  \bibfield  {author} {\bibinfo {author} {\bibfnamefont {C.~S.}\ \bibnamefont
  {Chiu}}, \bibinfo {author} {\bibfnamefont {G.}~\bibnamefont {Ji}}, \bibinfo
  {author} {\bibfnamefont {A.}~\bibnamefont {Mazurenko}}, \bibinfo {author}
  {\bibfnamefont {D.}~\bibnamefont {Greif}},\ and\ \bibinfo {author}
  {\bibfnamefont {M.}~\bibnamefont {Greiner}},\ }\href
  {https://doi.org/10.1103/PhysRevLett.120.243201} {\bibfield  {journal}
  {\bibinfo  {journal} {Physical Review Letters}\ }\textbf {\bibinfo {volume}
  {120}},\ \bibinfo {pages} {243201} (\bibinfo {year} {2018})}\BibitemShut
  {NoStop}%
\bibitem [{\citenamefont {Schreiber}\ \emph {et~al.}(2015)\citenamefont
  {Schreiber}, \citenamefont {Hodgman}, \citenamefont {Bordia}, \citenamefont
  {L{\"u}schen}, \citenamefont {Fischer}, \citenamefont {Vosk}, \citenamefont
  {Altman}, \citenamefont {Schneider},\ and\ \citenamefont
  {Bloch}}]{schreiber2015}%
  \BibitemOpen
  \bibfield  {author} {\bibinfo {author} {\bibfnamefont {M.}~\bibnamefont
  {Schreiber}}, \bibinfo {author} {\bibfnamefont {S.~S.}\ \bibnamefont
  {Hodgman}}, \bibinfo {author} {\bibfnamefont {P.}~\bibnamefont {Bordia}},
  \bibinfo {author} {\bibfnamefont {H.~P.}\ \bibnamefont {L{\"u}schen}},
  \bibinfo {author} {\bibfnamefont {M.~H.}\ \bibnamefont {Fischer}}, \bibinfo
  {author} {\bibfnamefont {R.}~\bibnamefont {Vosk}}, \bibinfo {author}
  {\bibfnamefont {E.}~\bibnamefont {Altman}}, \bibinfo {author} {\bibfnamefont
  {U.}~\bibnamefont {Schneider}},\ and\ \bibinfo {author} {\bibfnamefont
  {I.}~\bibnamefont {Bloch}},\ }\href {https://doi.org/10.1126/science.aaa7432}
  {\bibfield  {journal} {\bibinfo  {journal} {Science}\ }\textbf {\bibinfo
  {volume} {349}},\ \bibinfo {pages} {842} (\bibinfo {year}
  {2015})}\BibitemShut {NoStop}%
\bibitem [{\citenamefont {Choi}\ \emph {et~al.}(2016)\citenamefont {Choi},
  \citenamefont {Hild}, \citenamefont {Zeiher}, \citenamefont {Schau{\ss}},
  \citenamefont {{Rubio-Abadal}}, \citenamefont {Yefsah}, \citenamefont
  {Khemani}, \citenamefont {Huse}, \citenamefont {Bloch},\ and\ \citenamefont
  {Gross}}]{choi2016}%
  \BibitemOpen
  \bibfield  {author} {\bibinfo {author} {\bibfnamefont {J.-y.}\ \bibnamefont
  {Choi}}, \bibinfo {author} {\bibfnamefont {S.}~\bibnamefont {Hild}}, \bibinfo
  {author} {\bibfnamefont {J.}~\bibnamefont {Zeiher}}, \bibinfo {author}
  {\bibfnamefont {P.}~\bibnamefont {Schau{\ss}}}, \bibinfo {author}
  {\bibfnamefont {A.}~\bibnamefont {{Rubio-Abadal}}}, \bibinfo {author}
  {\bibfnamefont {T.}~\bibnamefont {Yefsah}}, \bibinfo {author} {\bibfnamefont
  {V.}~\bibnamefont {Khemani}}, \bibinfo {author} {\bibfnamefont {D.~A.}\
  \bibnamefont {Huse}}, \bibinfo {author} {\bibfnamefont {I.}~\bibnamefont
  {Bloch}},\ and\ \bibinfo {author} {\bibfnamefont {C.}~\bibnamefont {Gross}},\
  }\href {https://doi.org/10.1126/science.aaf8834} {\bibfield  {journal}
  {\bibinfo  {journal} {Science}\ }\textbf {\bibinfo {volume} {352}},\ \bibinfo
  {pages} {1547} (\bibinfo {year} {2016})}\BibitemShut {NoStop}%
\bibitem [{\citenamefont {Alet}\ and\ \citenamefont
  {Laflorencie}(2018)}]{alet2018}%
  \BibitemOpen
  \bibfield  {author} {\bibinfo {author} {\bibfnamefont {F.}~\bibnamefont
  {Alet}}\ and\ \bibinfo {author} {\bibfnamefont {N.}~\bibnamefont
  {Laflorencie}},\ }\href {https://doi.org/10.1016/j.crhy.2018.03.003}
  {\bibfield  {journal} {\bibinfo  {journal} {Comptes Rendus Physique}\
  }\bibinfo {series} {Quantum Simulation / {{Simulation}} Quantique},\ \textbf
  {\bibinfo {volume} {19}},\ \bibinfo {pages} {498} (\bibinfo {year}
  {2018})}\BibitemShut {NoStop}%
\bibitem [{\citenamefont {Abanin}\ \emph {et~al.}(2019)\citenamefont {Abanin},
  \citenamefont {Altman}, \citenamefont {Bloch},\ and\ \citenamefont
  {Serbyn}}]{abanin2019}%
  \BibitemOpen
  \bibfield  {author} {\bibinfo {author} {\bibfnamefont {D.~A.}\ \bibnamefont
  {Abanin}}, \bibinfo {author} {\bibfnamefont {E.}~\bibnamefont {Altman}},
  \bibinfo {author} {\bibfnamefont {I.}~\bibnamefont {Bloch}},\ and\ \bibinfo
  {author} {\bibfnamefont {M.}~\bibnamefont {Serbyn}},\ }\href
  {https://doi.org/10.1103/RevModPhys.91.021001} {\bibfield  {journal}
  {\bibinfo  {journal} {Reviews of Modern Physics}\ }\textbf {\bibinfo {volume}
  {91}},\ \bibinfo {pages} {021001} (\bibinfo {year} {2019})}\BibitemShut
  {NoStop}%
\bibitem [{\citenamefont {Yao}\ \emph {et~al.}(2024)\citenamefont {Yao},
  \citenamefont {Chi}, \citenamefont {Wang}, \citenamefont {Fletcher},\ and\
  \citenamefont {Zwierlein}}]{yao2024}%
  \BibitemOpen
  \bibfield  {author} {\bibinfo {author} {\bibfnamefont {R.}~\bibnamefont
  {Yao}}, \bibinfo {author} {\bibfnamefont {S.}~\bibnamefont {Chi}}, \bibinfo
  {author} {\bibfnamefont {M.}~\bibnamefont {Wang}}, \bibinfo {author}
  {\bibfnamefont {R.~J.}\ \bibnamefont {Fletcher}},\ and\ \bibinfo {author}
  {\bibfnamefont {M.}~\bibnamefont {Zwierlein}},\ }\href
  {https://doi.org/10.48550/arXiv.2411.08780} {\bibinfo {title} {Measuring pair
  correlations in {{Bose}} and {{Fermi}} gases via atom-resolved microscopy}}
  (\bibinfo {year} {2024}),\ \Eprint {https://arxiv.org/abs/2411.08780}
  {arXiv:2411.08780 [cond-mat]} \BibitemShut {NoStop}%
\bibitem [{\citenamefont {Xiang}\ \emph {et~al.}(2024)\citenamefont {Xiang},
  \citenamefont {{Cruz-Col{\'o}n}}, \citenamefont {Chua}, \citenamefont
  {Milner}, \citenamefont {de~Hond}, \citenamefont {Fricke},\ and\
  \citenamefont {Ketterle}}]{xiang2024}%
  \BibitemOpen
  \bibfield  {author} {\bibinfo {author} {\bibfnamefont {J.}~\bibnamefont
  {Xiang}}, \bibinfo {author} {\bibfnamefont {E.}~\bibnamefont
  {{Cruz-Col{\'o}n}}}, \bibinfo {author} {\bibfnamefont {C.~C.}\ \bibnamefont
  {Chua}}, \bibinfo {author} {\bibfnamefont {W.~R.}\ \bibnamefont {Milner}},
  \bibinfo {author} {\bibfnamefont {J.}~\bibnamefont {de~Hond}}, \bibinfo
  {author} {\bibfnamefont {J.~F.}\ \bibnamefont {Fricke}},\ and\ \bibinfo
  {author} {\bibfnamefont {W.}~\bibnamefont {Ketterle}},\ }\href
  {https://doi.org/10.48550/arXiv.2411.08779} {\bibinfo {title} {In situ
  imaging of the thermal de {{Broglie}} wavelength in an ultracold {{Bose}}
  gas}} (\bibinfo {year} {2024}),\ \Eprint {https://arxiv.org/abs/2411.08779}
  {arXiv:2411.08779 [cond-mat]} \BibitemShut {NoStop}%
\bibitem [{\citenamefont {Jin}\ \emph {et~al.}(2024)\citenamefont {Jin},
  \citenamefont {Dai}, \citenamefont {Verstraten}, \citenamefont {Dixmerias},
  \citenamefont {Alhyder}, \citenamefont {Salomon}, \citenamefont {Peaudecerf},
  \citenamefont {De~Jongh},\ and\ \citenamefont {Yefsah}}]{jin2024}%
  \BibitemOpen
  \bibfield  {author} {\bibinfo {author} {\bibfnamefont {S.}~\bibnamefont
  {Jin}}, \bibinfo {author} {\bibfnamefont {K.}~\bibnamefont {Dai}}, \bibinfo
  {author} {\bibfnamefont {J.}~\bibnamefont {Verstraten}}, \bibinfo {author}
  {\bibfnamefont {M.}~\bibnamefont {Dixmerias}}, \bibinfo {author}
  {\bibfnamefont {R.}~\bibnamefont {Alhyder}}, \bibinfo {author} {\bibfnamefont
  {C.}~\bibnamefont {Salomon}}, \bibinfo {author} {\bibfnamefont
  {B.}~\bibnamefont {Peaudecerf}}, \bibinfo {author} {\bibfnamefont
  {T.}~\bibnamefont {De~Jongh}},\ and\ \bibinfo {author} {\bibfnamefont
  {T.}~\bibnamefont {Yefsah}},\ }\href
  {https://doi.org/10.1103/PhysRevResearch.6.013158} {\bibfield  {journal}
  {\bibinfo  {journal} {Physical Review Research}\ }\textbf {\bibinfo {volume}
  {6}},\ \bibinfo {pages} {013158} (\bibinfo {year} {2024})}\BibitemShut
  {NoStop}%
\end{thebibliography}
\end{document}